\documentclass[AMA,STIX1COL]{WileyNJD-v2}
\usepackage{enumitem}
\usepackage{bm}
\usepackage{graphicx}
\usepackage{algorithm}
\usepackage{color}
\usepackage{moreverb}
\usepackage{subcaption}
\usepackage{caption} 
\usepackage{tikz}
\usetikzlibrary{patterns}
\usepackage{pgfplots}
\pgfplotsset{compat=1.8}
\usetikzlibrary{pgfplots.groupplots}
\usepackage{pgfplotstable, booktabs}
\usepackage{tablefootnote}

\newcommand{\func}[3]{\ensuremath{#1 : #2 \rightarrow #3}}

\newcommand{\Acal}{\ensuremath{\mathscr{A}}}
\newcommand{\Bcal}{\ensuremath{\mathscr{B}}}

\newcommand{\Scal}{\ensuremath{\mathscr{S}}}

\newcommand{\Vcal}{\ensuremath{\mathscr{V}}}

\newcommand{\Rbb}{\ensuremath{\mathbb{R} }}

\newcommand\Bbm{{\ensuremath{\bm{B}}}}

\newcommand\Ebm{{\ensuremath{\bm{E}}}}
\newcommand\Fbm{{\ensuremath{\bm{F}}}}
\newcommand\Gbm{{\ensuremath{\bm{G}}}}

\newcommand\Ibm{{\ensuremath{\bm{I}}}}

\newcommand\Kbm{{\ensuremath{\bm{K}}}}

\newcommand\Mbm{{\ensuremath{\bm{M}}}}
\newcommand\Nbm{{\ensuremath{\bm{N}}}}

\newcommand\Pbm{{\ensuremath{\bm{P}}}}
\newcommand\Qbm{{\ensuremath{\bm{Q}}}}
\newcommand\Rbm{{\ensuremath{\bm{R}}}}
\newcommand\Sbm{{\ensuremath{\bm{S}}}}
\newcommand\Tbm{{\ensuremath{\bm{T}}}}
\newcommand\Ubm{{\ensuremath{\bm{U}}}}
\newcommand\Vbm{{\ensuremath{\bm{V}}}}
\newcommand\Wbm{{\ensuremath{\bm{W}}}}
\newcommand\Xbm{{\ensuremath{\bm{X}}}}
\newcommand\Ybm{{\ensuremath{\bm{Y}}}}

\newcommand\fbm{{\ensuremath{\bm{f}}}}
\newcommand\gbm{{\ensuremath{\bm{g}}}}

\newcommand\nbm{{\ensuremath{\bm{n}}}}

\newcommand\ubm{{\ensuremath{\bm{u}}}}

\newcommand\wbm{{\ensuremath{\bm{w}}}}
\newcommand\xbm{{\ensuremath{\bm{x}}}}
\newcommand\ybm{{\ensuremath{\bm{y}}}}
\newcommand\zbm{{\ensuremath{\bm{z}}}}

\newcommand\Cbold{\ensuremath{\mathbf{C}}}

\newcommand\Ibold{\ensuremath{\mathbf{I}}}

\newcommand\vbold{\ensuremath{\mathbf{v}}}

\newcommand\lambdabold{{\ensuremath{\boldsymbol{\lambda}}}}

\newcommand\varphibold{{\ensuremath{\boldsymbol{\varphi}}}}

\newcommand\Pibold{{\ensuremath{\boldsymbol{\Pi}}}}

\articletype{Research Article}

\received{}
\revised{}
\accepted{}

\begin{document}

\title{A Computationally Tractable Framework for Nonlinear Dynamic Multiscale Modeling of Membrane Woven Fabrics}

\author[1]{Philip Avery}
\author[3]{Daniel Z. Huang}
\author[2]{Wanli He}
\author[2]{Johanna Ehlers}
\author[4]{Armen Derkevorkian}
\author[1,2,3]{Charbel Farhat}

\authormark{AVERY \textsc{et al}}

\address[1]{\orgdiv{Department of Aeronautics and Astronautics}, \orgname{Stanford University},
            \orgaddress{Stanford, \state{CA} 94305, \country{USA}}}
\address[2]{\orgdiv{Department of Mechanical Engineering}, \orgname{Stanford University},
            \orgaddress{Stanford, \state{CA} 94305, \country{USA}}}
\address[3]{\orgdiv{Institute for Computational and Mathematical Engineering}, \orgname{Stanford University},
            \orgaddress{Stanford, \state{CA} 94305, \country{USA}}}
\address[4]{\orgdiv{Jet Propulsion Laboratory}, \orgname{California Institute of Technology},
            \orgaddress{Pasadena, \state{CA} 91109, \country{USA}}}

\abstract[Summary]{\normalsize
A general-purpose computational homogenization framework is proposed for the nonlinear dynamic analysis
of membranes exhibiting complex microscale and/or mesoscale heterogeneity characterized by in-plane periodicity
that cannot be effectively treated by a conventional method, such as woven fabrics. The framework is
a generalization of the ``finite element squared'' (or FE$^2$) method in which a localized portion of
the periodic subscale structure is modeled using finite elements. The numerical solution of displacement
driven problems involving this model can be adapted to the context of membranes by a variant of the Klinkel-Govindjee
method \cite{klinkel2002using} originally proposed for using finite strain, three-dimensional material
models in beam and shell elements. This approach relies on numerical enforcement of the plane stress
constraint and is enabled by the principle of frame invariance. Computational tractability is achieved
by introducing a regression-based surrogate model informed by a physics-inspired training regimen in
which FE$^2$ is utilized to simulate a variety of numerical experiments including uniaxial, biaxial and
shear straining of a material coupon. Several alternative surrogate models are evaluated including an
artificial neural network. The framework is demonstrated and validated for a realistic Mars landing application
involving supersonic inflation of a parachute canopy made of woven fabric.
}

\maketitle  


\section{Introduction}

\begin{figure}[ht]
\centering
\includegraphics[width=0.6\linewidth]{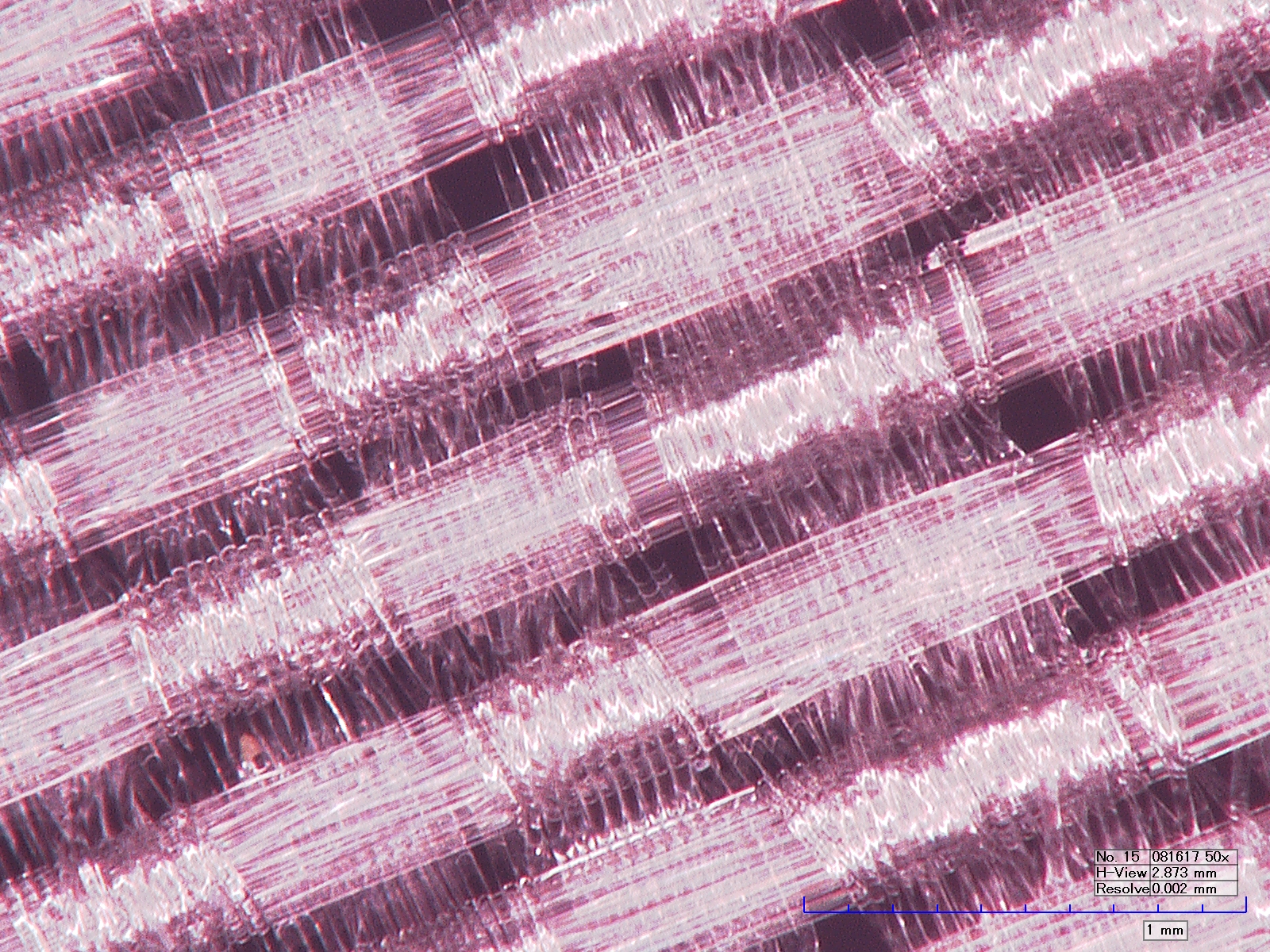}
	\caption{Optical microscope imaging of yarns of a parachute membrane made of a woven fabric under tensile loading (image courtesy of NASA/JPL-Caltech).}
\label{fig:figure1}
\end{figure}

Nonlinear multiscale problems -- defined here as nonlinear problems exhibiting vastly different
scale features that are significant to the macroscopic behavior -- are ubiquitous in science and engineering.
They arise, for example, in the modeling of woven fabrics (see Figure \ref{fig:figure1}) used in body
armor and inflatable structures such as vehicle air bags, parachutes and other atmospheric decelerators;
and in the modeling of textiles within the context of forming processes for woven composites \cite{gereke2013experimental}.
Numerical methods that attempt to resolve all relevant scales typically lead to massive discretized problems.
However, recent developments using a variety of alternative surrogate modeling techniques -- including
nonlinear, projection-based model order reduction \cite{zahr2017multilevel, yvonnet2007reduced, he2020situ}, kriging \cite{knap2008adaptive},
and artificial neural networks (NNs) \cite{le2015computational, lu2019data} -- to accelerate the solution of one or more
scales within the context of a computational homogenization framework present a coherent methodology
by which a computationally tractable approximation can be attained without resorting to ad-hoc approximations.
Notably, thin shell and membrane discretizations have not been considered in this context prior to this
work, although several frameworks for multiscale modeling of shells without emphasis on computational
efficiency have been proposed \cite{coenen2010computational, ha2011numerical, larsson2013stress}. In
particular, this paper addresses the case of a \emph{hybrid} discretization in which plane stress membrane
elements are employed at the macroscopic scale, for the sake of convenience and numerical efficiency; but
three-dimensional (3D) solid elements are preferred at the mesoscopic and/or microscopic scales for the sake of generality and in order to most
precisely represent geometric features and deformation modes at these scales.

In this paper, the presented multiscale framework is built on a macro-micro concept that generalizes to $n$-levels, although without
loss of generality, only the two-level case is presented, anticipating that this case is sufficient for many problems of interest. The framework allows for the treatment of unilateral contact constraints
at both macroscale and microscale. It features a microscale model discretized with solid elements, to allow accurate representation of microscopic geometric features such as yarns and voids. However,
the proposed framework readily generalizes to alternative microscale discretizations such as shell elements.
For the case of a macroscale model that is also discretized with solid elements, well-established localization/homogenization
scale bridging strategies have been developed \cite{miehe1999computational} and typically provide a mapping
between the 3D deformation gradient ($\Fbm$) and first Piola-Kirchhoff stress ($\Pbm$)
tensors from which a constitutive relation is inferred. However, when the macroscale model is discretized
with membrane elements and the microscale is discretized with solid elements, coupling between the
two scales requires careful attention. The treatment proposed in this paper has two novel components:
\begin{itemize}[topsep=0pt] 
\itemsep-0.35em
\item First, it is observed that due to the principle of material frame invariance, the conventional 3D
      $\Fbm$--$\Pbm$ scale bridging approach can be reformulated using the polar decomposition of the
      deformation gradient to furnish a mapping between the right stretch tensor and its conjugate, the
      symmetric Biot stress tensor. Using this reformulation combined with some straightforward transformations,
      it is shown that an unconventional FE$^2$ material model can \emph{conveniently} be used as a
      drop-in replacement for any conventional 3D material model formulated as a mapping
      between the commonly used Green-Lagrange strain ($\Ebm$) and its conjugate, the second Piola-Kirchhoff
      stress tensor ($\Sbm$). The relevance of this development to the issue of multiscale membrane-solid
      coupling will be addressed in what follows.
\item Material models used in membrane elements are typically of the plane stress variety. In some cases,
      a plane stress variant of a 3D material model for which the plane stress condition is enforced analytically can be derived. 
      When an analytical solution is not available, numerical
      enforcement of the plane stress condition is commonly used, for example, in the case of J2 elastoplasticity
      \cite{simo1988exact}. This involves solving numerically -- using a root-finding method such as Newton's
      method or the bisection method -- a nonlinear equation to enforce the plane stress condition. Klinkel
      and Govindjee have shown \cite{klinkel2002using} how numerical enforcement of the plane stress
      condition can be used to construct an interface that, in principle, enables any 3D
      material model to be ``converted'' into a plane stress variant that can then be used in a shell
      or beam element. Because the plane stress condition is typically expressed by constraints on the
      out-of-plane components of the \emph{second Piola-Kirchhoff} stress tensor, the method of Klinkel
      and Govindjee was presented in the most convenient way -- that is, using the $\Ebm$--$\Sbm$ conjugate
      pair. Here, it is simply noted that this method can be trivially adapted to membranes and furthermore can be 
      conveniently used with any material model of the form $\Sbm = \widehat{\Sbm}(\Ebm)$ including, but not limited to, constitutive relations inferred from a FE$^2$ computational 
      homogenization formulated using polar decomposition as outlined above.
\end{itemize}
Motivated by the fact that the proposed plane stress constitutive law is essentially a mapping is between
two pairs of 3D vectors, a lightweight alternative is considered, in which a regression based-model is used as a surrogate for constitutive function evaluations that would otherwise 
require the solution of a finite-element model of the microscale RVE. Three alternative surrogates, each capable of achieving computational
tractability, are presented and evaluated: (1) the classical linear elastic model fitted to data using
linear regression; (2) a quadratic model fitted to data using linear regression; and (3) an artificial
NN model fitted to data using the PyTorch library. In each case, the data used to train and test
the model is obtained by exercising the proposed high-fidelity multiscale membrane model on a series
of numerical experiments intended to mimic the familiar physical experimental-based methodology typically
used in the development of conventional material models. A novel, nonlinear, projection-based model order reduction (PMOR) approach is also
proposed for accelerating the training process and described in Appendix \ref{sec:micro-rom}.

The remainder of this paper is organized as follows. Section \ref{sec:multi-hdm} provides an overview
of the proposed two-level multiscale framework with a locally attached microscale, focusing on the context
of large-deformation structural mechanics with macroscopic discretization using membrane elements, microscale
discretization using 3D solid elements, bridging between the scales and the solution of the discrete
coupled multiscale problem including treatment of contact at both scales. In Section \ref{sec:micro-surrogate},
three regression-based surrogate microscale models and their training methodologies are presented and
compared. A numerical example is provided in Section \ref{sec:app} to evaluate the proposed framework, involving
a realistic, coupled, multiscale fluid-structure simulation of the deployment of a Disk-Gap-Band (DGB) parachute in the Martian atmosphere.
Finally, conclusions are offered in Section \ref{sec:conclude}.

\section{Multiscale Formulation for Membranes Based on a Locally Attached Microstructure}
\label{sec:multi-hdm}

In this section, a multiscale continuum mechanics formulation suitable for membranes and based on the concept
of a locally attached microstructure is presented. As formulated, the stress-strain relationship for
a heterogeneous membrane is not defined by a conventional plane stress constitutive law, but rather by:
(a) the solution at each material point of one or more boundary value problems governing its microstructure;
and (b) the numerical enforcement of the plane stress condition. Although the concept generalizes naturally
to three or more scales, it is presented and applied here to problems that exhibit precisely two separate scales --
for the sake of clarity. Specifically, the stress-strain relationship at the coarse scale is defined by the
solution of boundary value problems at the fine scale, an appropriate scale transition method and
a constraint on the out-of-plane components of the homogenized stress tensor. At the finest scale, where
all heterogeneities can be adequately resolved and described by an available constitutive theory, it
is defined by an analytical constitutive law. All considered length scales are assumed to be much larger
than the molecular dimension so that the continuum assumption holds. Furthermore, scale separation is
assumed to loosely couple the various scales through localization from coarse to fine scales and homogenization
from fine to coarse scales. For further details, the reader is referred to \cite{smit1998prediction,
miehe1999computational, feyel2000fe2, kouznetsova2001approach} for the concept of a locally attached
microstructure. The approach adopted here can be interpreted as a generalization and/or application
to the case of membranes of the localization/homogenization scale bridging strategy presented in \cite{miehe1999computational}.

\subsection{Preliminaries}
\label{subsec:prelim}

Consider a domain $\Bcal_0 \subset \Rbb^3$ defining a highly heterogeneous membrane structure of
interest.  Assume that its boundary $\partial\Bcal_0$ is subject to prescribed displacements on
$\partial\Bcal_0^\ubm \subset \partial\Bcal_0$ and tractions on $\partial\Bcal_0 \setminus \partial\Bcal_0^\ubm$.
Let $\func{\varphibold_0^t}{\Bcal_0}{\Bcal_0^t}$ denote the nonlinear transformation that maps a point
in the reference configuration, $\Xbm_0 \in \Bcal_0$, at time $t$, to a counterpart in the current
configuration, $\xbm_0(\Xbm_0; t) = \varphibold_0^t(\Xbm_0) \in \Bcal_0^t$. Here, the current configuration
of the membrane $\Bcal_0^t$ is defined as
\begin{equation*}
  \Bcal_0^t = \left \{ \xbm_0 \in \Rbb^3 \: | \: \xbm_0 = \phi\left(\xi^{(1)}, \xi^{(2)}\right) + \xi^{(3)}\nbm_0 \right \}
\end{equation*}
where the map $\phi : \Acal_0 \rightarrow \Rbb^3$ defines the current position of the mid-surface of
the membrane, $\left(\xi^{(1)}, \xi^{(2)}\right) \in \Acal_0 \subset \Rbb^2$ are coordinates parameterizing the
mid-surface, $\nbm_0 \in \Rbb^3$ is the unit normal to the mid-surface in the current configuration,
$\xi^{(3)} \in \left[ -h/2, h/2 \right]$ is a coordinate parameterizing the direction normal to the surface
and $h$ is the upper bound of the membrane thickness. Similarly, the reference configuration of the membrane
$\Bcal_0$ is defined as
\begin{equation*}
  \Bcal_0 = \left\{ \Xbm_0 \in \Rbb^3 \: | \: \Xbm_0 = \Phi\left(\xi^{(1)}, \xi^{(2)}\right) + \xi^{(3)}\Nbm_0 \right\}
\end{equation*}
where the map $\Phi : \Acal_0 \rightarrow \Rbb^3$ defines the reference position of the mid-surface
of the membrane and $\Nbm_0 \in \Rbb^3$ is the unit normal to the mid-surface in the reference configuration.
The deformation of this domain is governed by a reduction of the finite deformation continuum equations
to the mid-surface with a plane stress but otherwise unknown constitutive law due to the assumed highly
heterogeneous fine scale structure. For this reason, generalizing the work described in
\cite{smit1998prediction, miehe1999computational, feyel2000fe2, kouznetsova2001approach}, 
the deformation problem is solved here by locally attaching an appropriately defined microstructure to
each mid-surface point, computing the stress-strain relationship at each such point through the solution
of a microstructure boundary value problem, bridging the scales via a localization and homogenization
strategy and numerically enforcing the plane stress constraint on the resulting homogenized stress tensor.
An appropriately defined microstructure in this context is one that represents only a minuscule
``representative surface element'' (RSE) of the membrane within which the entire thickness of the membrane
is accounted for. Hence, the range of the in-plane coordinates $\xi^{(1)}, \xi^{(2)}$ in
the microscale domain should be much smaller than in the macroscale domain, while the ranges of the normal
coordinate $\xi^{(3)}$ should be identical in both domains. The separation of scales and assumed periodicity
in only two of the three spatial dimensions are notable characteristics of the problem of interest and
its proposed treatment that distinguish it from the ubiquitous alternative multiscale treatments devised
for fully 3D scale bridging.

Here and throughout the remainder of this paper, the subscripts $0$ and $1$ denote quantities associated
with the coarse ($0$-th) and fine ($1$-st) scales, respectively. For simplicity, an $l$-th scale is also
referred to as scale $l$ or level $l$, interchangeably. The deformation at both scales is governed by
the finite deformation continuum equations, with the stress-strain relationship defined by the solution
of a constrained boundary value problem formulated at a finer scale for level $l = 0$, or an assumed constitutive
law at the fine scale designated by level $l = 1$. Let $\func{\varphibold_1}{\Bcal_1}{\Bcal_1^{\prime}}$
denote the nonlinear transformation that maps a point in the fine scale reference configuration, $\Xbm_1 \in \Bcal_1$,
to a counterpart in the fine scale current configuration, $\xbm_1(\Xbm_1) = \varphibold_1(\Xbm_1) \in \Bcal_1^{\prime}$. 
As in the formulation of the macroscale problem, $\partial\Bcal_1$ is defined as the boundary of $\Bcal_1$
and $\partial\Bcal_1^\ubm$ as its part where a displacement is prescribed.

The boundary conditions at scale $0$ are defined by the physical problem of interest, while those at scale $1$ depend on the deformations at the coarse scale. The constitutive
law at the fine scale is chosen based on the expected response of this scale, while there is no preassigned constitutive law at the coarse scale but rather a dependence on the response of
the microstructure to evaluate the constitutive function. Arbitrarily complex fine scale constitutive relationships involving nonlinearities and path-dependency are
allowed, although in what follows only constitutive functions described by
\begin{equation*}
  \Sbm_1 = \widehat{\Sbm}_1\left(\Ebm_1\right)
\end{equation*}
are considered, where $\Sbm_1$ and $\Ebm_1$ denote the microscale second Piola-Kirchhoff stress tensor and
Green-Lagrange strain tensor, respectively, and $\widehat{\Sbm}_1$ is the microscale
constitutive function. At the coarse scale, the intent is to devise a multiscale, plane stress constitutive function
of the form
\begin{equation}
\label{eqn:claw0}
  \Sbm_0^m = {\widehat{\Sbm}}_0^m\left(\Ebm_0^m\right)
\end{equation}
where $\Sbm_0$ and $\Ebm_0$ denote the macroscale second Piola-Kirchhoff stress tensor and Green-Lagrange
strain tensor, respectively; and the superscript $m$ applied to a tensor quantity designates the restriction of the tensor to its in-plane
membrane components. For example, the membrane part of $\Sbm_0$ is given by
\begin{equation*}
	\Sbm_0^m = \begin{bmatrix} S_0^{(11)} & S_0^{(12)} \\[1ex]
                             S_0^{(21)} & S_0^{(22)} \end{bmatrix}
\end{equation*}
The superscript $m$ applied to a \emph{constitutive function} (for example ${\widehat{\Sbm}}_0^m$) indicates
that the function is a particular plane stress type of constitutive relation that evaluates the in-plane
membrane components of a stress tensor while constraining its out-of-plane components to be identically zero. A general
numerical procedure for constructing such a function will be described subsequently.

\subsection{Scale bridging}
\label{subsec:sb}

Following the work presented in \cite{miehe1999computational}, the boundary conditions on $\Bcal_1$
are defined so that the pointwise deformation gradient tensor at level $0$, $\Fbm_0$,
is equal to the volumetric average of the deformation gradient tensor at level $1$ -- that is,
\begin{equation*}
  \Fbm_0 = \frac{1}{|\Bcal_{1}|}\int_{\Bcal_{1}} \Fbm_{1}\,dV
\end{equation*}
This \emph{localization} transmission condition can be conveniently enforced by prescribing a boundary
deformation of the form
\begin{equation}
\label{eqn:micro-deform-ansatz}
  \left.\xbm_1\right|_{\partial\Bcal_1^\ubm} = \left.\Xbm_1\right|_{\partial\Bcal_1^\ubm} \Fbm_0 + \wbm_1
\end{equation}
subject to some conditions (see \cite{zahr2017multilevel}), where $\wbm_1$ represents the non-uniform
part of the boundary deformation. Without loss of generality, the uniform essential boundary condition $\wbm_1 = 0$ is assumed.

The pointwise first Piola-Kirchhoff stress tensor at level $0$ is defined as the volumetric average of
the stress tensor at level $1$
\begin{equation}
\label{eqn:micro-macro}
  \Pbm_0 = \frac{1}{|\Bcal_1|}\int_{\Bcal_1} \Pbm_1\,dV
\end{equation}
This \emph{homogenization} transmission-type condition can be conveniently determined from quantities defined
solely on $\partial\Bcal_1^\ubm$ by applying a Gauss-type identity to (\ref{eqn:micro-macro})
\begin{subequations}
\label{eqn:micro-macro-2}
\begin{align}
  \Pbm_0 &= \frac{1}{|\Bcal_1|}\int_{\partial\Bcal_1^\ubm} \Pbm_1\Nbm_1 \otimes \Xbm_1\,dA \\
         &= \frac{1}{|\Bcal_1|} \left.\Xbm_1^T\right|_{\partial\Bcal_1^\ubm} \left.\fbm_1\right|_{\partial\Bcal_1^\ubm}
\end{align}
\end{subequations}
where $\left.\fbm_1\right|_{\partial\Bcal_1^\ubm}$ is the vector of so-called \emph{reaction forces}
associated with the prescribed deformations (\ref{eqn:micro-deform-ansatz}) and the superscript $T$
designates the transpose operation.

\begin{figure}[ht]
\centering
\includegraphics[width=0.8\linewidth]{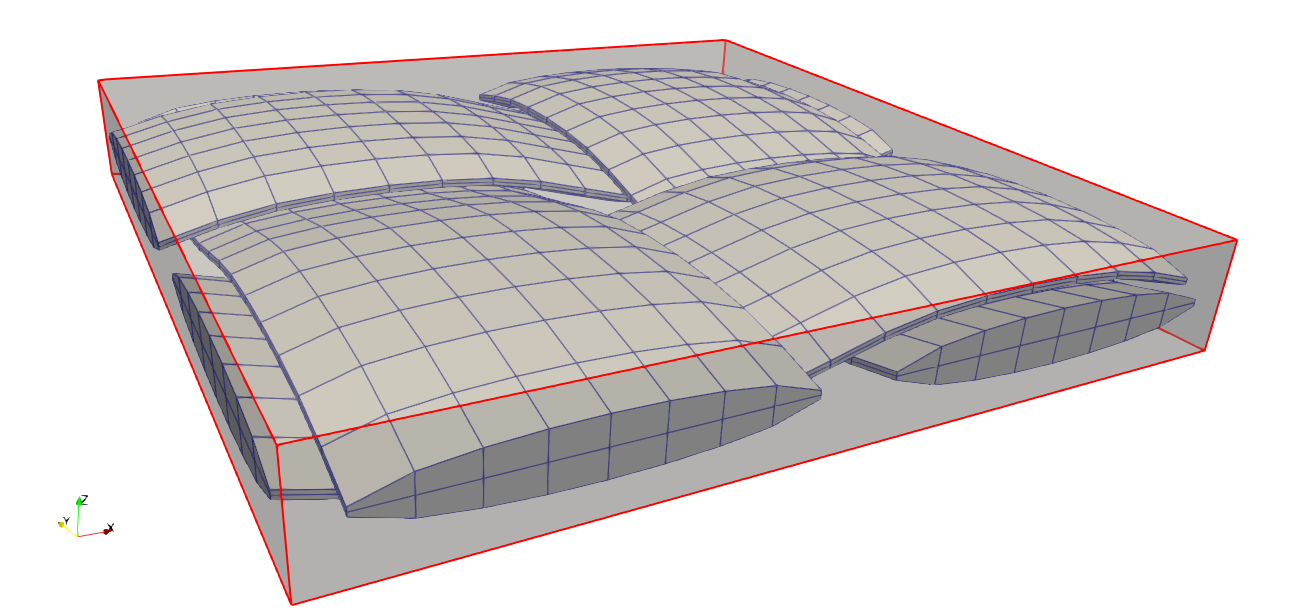}
	\caption{Representative Surface Element of a membrane within which the membrane's thickness is represented in its entirety.}
\label{fig:figure2}
\end{figure}

In this context, the microscale volume measure $|\Bcal_1|$ should be interpreted as the entire volume of
a bounding box enclosing the microscale volume (see Figure \ref{fig:figure2}), including both regions of
solid material and voids. The height of the bounding box $h+\varepsilon$ should be slightly larger than the minimum enclosing
dimension $h$ in the $\xi^{(3)}$ direction (i.e. $\varepsilon > 0$) so that the microscale volume $\Bcal_1$
does not intersect the box's upper and lower faces. The magnitude of $\varepsilon$ is otherwise arbitrary,
as the dependence of the homogenized stress tensor on this parameter will be subsequently canceled when evaluating the membrane stress resultant. 
Note that the boundary ${\partial\Bcal_1^\ubm}$ used to define the transmission-type conditions is entirely contained within the four side faces of the bounding box, 
i.e., the faces whose normals coincide with the $\xi^{(1)}$ and $\xi^{(2)}$ axes.

Equations (\ref{eqn:micro-deform-ansatz}) and (\ref{eqn:micro-macro-2}) constitute a relation of the form
\begin{equation}
\label{eqn:claw0PF}
  \Pbm_0 = \widehat{\Pbm}_0\left(\Fbm_0\right)
\end{equation}
that is evaluated in three steps as follows:
\begin{enumerate}[topsep=0pt]
\itemsep-0.35em
\item First, the microscale problem defined with the boundary conditions given in (\ref{eqn:micro-deform-ansatz})
      is solved.
\item Next, the solution of the microscale problem is postprocessed to obtain the reaction forces. This can be done either by a boundary integral as shown in (\ref{eqn:micro-macro-2}a), or 
	by a volume integral over the region of the domain adjacent to the boundary \cite{melbo2003goal}.  
\item Finally, the reaction forces are combined and scaled according to (\ref{eqn:micro-macro-2}b) to
      produce the homogenized first Piola-Kirchhoff stress tensor $\Pbm_0$.
\end{enumerate}

Unfortunately, (\ref{eqn:claw0PF}) is not directly compatible with the stated application of interest -- namely, a \emph{plane stress} relation of the form (\ref{eqn:claw0}) expressed in terms of 
the in-plane components of the Green-Lagrange strain and second Piola-Kirchhoff stress tensors. To formally adapt the homogenization methodology to this setting, it is first assumed without loss of 
generality that the relation (\ref{eqn:claw0PF}) satisfies the principle of material frame invariance, which can be stated as follows \cite{ciarlet1988mathematical}
\begin{equation}
\label{eqn:frameinv}
  \widehat{\Pbm}_0\left(\Qbm \Fbm_0\right) = \Qbm \widehat{\Pbm}_0\left(\Fbm_0\right), \quad \forall \: \Qbm \in SO(3)
\end{equation}
where $SO(3)$ is the group of special orthogonal transformations defined as
\begin{equation*}
  SO(3) = \left\{ \Qbm \in \Rbb^3 : \Qbm^T\Qbm = \Qbm\Qbm^T = \Ibm, \: \det(\Qbm) = 1 \right\}
\end{equation*}
and $\Ibold$ is the identity matrix of dimension three. Regarding the assumption of material frame indifference, it can be shown under some conditions
that (\ref{eqn:claw0PF}) is objective \cite{yvonnet2013computational} and hence (\ref{eqn:frameinv}) holds. For cases where (\ref{eqn:frameinv}) does not hold, the alternative formulation 
proposed next can be interpreted as imposing or restoring material frame invariance, which is generally considered to be appropriate for constitutive relations in solid mechanics.

From the polar decomposition of the deformation gradient
\begin{equation*}
  \Fbm_0 = \Rbm_0 \Ubm_0
\end{equation*}
where $\Rbm_0 \in SO(3)$ is the rotation tensor and $\Ubm_0 \in \Rbb^{3\times 3}$ is the symmetric positive definite right stretch tensor, and the following expression of the unsymmetric Biot stress 
tensor 
\begin{equation*}
  \Bbm_0 = \Rbm_0^T\Pbm_0
\end{equation*}
it follows from (\ref{eqn:frameinv}) with $\Qbm = \Rbm_0^T$ that the homogenized constitutive law (\ref{eqn:claw0PF}) can equivalently be stated as a relation between the right stretch tensor and the
unsymmetric Biot stress using the same functional form, i.e.
\begin{equation}
\label{eqn:claw0BU}
  \Bbm_0 = \widehat{\Pbm}_0\left(\Ubm_0\right)
\end{equation}
This result can be simply interpreted as a variant of the standard transmission-type conditions (\ref{eqn:micro-deform-ansatz}, \ref{eqn:micro-macro-2}) in which the right stretch tensor is used 
instead of the deformation gradient to compute the microscale prescribed boundary deformations and the homogenized stress tensor obtained
by evaluating the constitutive function $\widehat{\Pbm}_0$ is identified as the Biot measure rather than
the first Piola-Kirchhoff. Specifically,
\begin{subequations}
\label{eqn:vartrans}
\begin{align}
  \left.\xbm_1\right|_{\partial\Bcal_1^\ubm} &= \left.\Xbm_1\right|_{\partial\Bcal_1^\ubm} \Ubm_0 \\
  \Bbm_0 &= \frac{1}{|\Bcal_1|} \left.\Xbm_1^T\right|_{\partial\Bcal_1^\ubm} \left.\fbm_1\right|_{\partial\Bcal_1^\ubm}
\end{align}
\end{subequations}

A more convenient relation between the Green-Lagrange strain and the first Piola-Kirchhoff stress tensor can
be obtained from (\ref{eqn:claw0BU}) by applying well-known transformations \cite{ogden2003nonlinear} as follows:
\begin{itemize} [topsep=0pt] 
\itemsep-0.35em
\item
First, the right stretch tensor can be obtained from the Green-Lagrange strain using
\begin{equation}
\label{eqn:transEU}
  \Ubm_0 = (\Cbold_0)^\frac{1}{2} = \sum_{i=1}^{i=3} \lambda_i N_i \otimes N_i
\end{equation}
where $\Cbold_0 = 2\Ebm_0 + \Ibold$ is the right Cauchy-Green deformation tensor and $\lambda_i^2$
and $N_i$ are the eigenvalues and eigenvectors, respectively, of $\Cbold_0$.
\item
Second, the second Piola-Kirchhoff stress can be obtained from the Biot stress using
\begin{equation}
\label{eqn:transBS}
  0.5\left(\Sbm_0 \Ubm_0 + \Ubm_0 \Sbm_0\right) = \Tbm_0
\end{equation}
where $\Tbm_0$ is the symmetric part of the Biot stress tensor $\Tbm_0 = 0.5\left(\Bbm_0 + \Bbm_0^T\right)$.
Note that (\ref{eqn:transBS}) has the form of the Lyapunov equation whose solution is given by a linear
system of equations, namely
\begin{equation}
\label{eqn:transBS2}
  \mathrm{vec}\left(\Sbm_0\right) = \left[\Ibold \otimes \Ubm_0 + \Ubm_0^T \otimes \Ibm\right]^{-1} \mathrm{vec}\left(2\Tbm_0\right)
\end{equation}
where $\otimes$ denotes the Kronecker product and $\mathrm{vec}\left(\cdot\right)$ denotes vectorization.
For example, the vectorization of $\Sbm_0$ is given by
\begin{equation*}
  \mathrm{vec}\left(\Sbm_0\right) =
  \begin{bmatrix} S_0^{(11)} & S_0^{(21)} & S_0^{(31)} & S_0^{(12)} & S_0^{(22)} & S_0^{(32)} & S_0^{(13)} & S_0^{(23)} & S_0^{(33)} \end{bmatrix}^T
\end{equation*}
Due to symmetry, the dimension of (\ref{eqn:transBS2}) can be further reduced to six.
\end{itemize}

Substituting (\ref{eqn:transEU}) and (\ref{eqn:transBS2}) into (\ref{eqn:claw0BU}) produces a constitutive
function relating the macroscale Green-Lagrange strain and second Piola-Kirchhoff stress tensor of the form
\begin{equation}
\label{eqn:claw0SE}
  \Sbm_0 = \widehat{\Sbm}_0\left(\Ebm_0\right)
\end{equation}
that is evaluated in five steps as follows:
\begin{enumerate}[topsep=0pt]
\itemsep-0.35em
\item First, the macroscale right stretch tensor $\Ubm_0$ is computed from the Green-Lagrange strain
      $\Ebm_0$.
\item Second, the microscale problem with prescribed boundary values given by (\ref{eqn:vartrans}a) is
      solved.
\item Third, the solution of the microscale problem is postprocessed to obtain the reaction forces.
\item Next, the reaction forces are combined and scaled according to (\ref{eqn:vartrans}b) to produce
      the homogenized unsymmetric Biot stress tensor $\Bbm_0$.
\item Finally, the Lyapunov equation (\ref{eqn:transBS2}) is solved to get the homogenized second Piola-Kirchhoff stress tensor $\Sbm_0$.
\end{enumerate}

\subsection{Three-dimensional finite strain material models for membrane elements}
\label{subsec:using}

The 3D constitutive law (\ref{eqn:claw0SE}) can be adapted to plane stress (and
hence membrane elements) using a variant of the method proposed by Klinkel and Govindjee \cite{klinkel2002using}
for using finite strain 3D material models in beam and shell elements, which in turn is closely related to earlier methods proposed
by De Borst \cite{de1991zero} and Dvorkin et al \cite{dvorkin1995formulation}. This method involves
solving a local nonlinear equation using Newton's method to enforce the plane stress condition. Specifically,
the requirement that the out-of-plane components of the second Piola-Kirchhoff stress tensor are zero, i.e.
\begin{equation}
\label{eqn:sz}
  \Sbm_0^z = \begin{bmatrix} S_0^{(33)} & S_0^{(13)} & S_0^{(31)} \end{bmatrix}^T = 0
\end{equation}
is enforced by iteratively solving for the corresponding out-of-plane components $\Ebm_0^z$ of the Green-Lagrange strain tensor, which are treated as unknowns
in the above equations. Each Newton iteration incurs a single evaluation of the 3D constitutive function (\ref{eqn:claw0SE}) and its constitutive tangent.
Solving the plane stress equation (\ref{eqn:sz}) for $\Ebm_0^z$ given $\Ebm_0^m$, then evaluating the in-plane components of the second Piola-Kirchhoff stress tensor $\Pbm_0^m$ at the resulting 
configuration corresponds to evaluating a plane stress constitutive relation of the form (\ref{eqn:claw0}), which can be used as a drop-in replacement for a conventional finite strain plane stress 
constitutive equation. This will be demonstrated in what follows using the general purpose finite element analyzer AERO-S \cite{geuzaine2003aeroelastic, farhat2003application}. 
\begin{remark}
The enforcement of the plane stress condition can be done using alternative conjugate pairings, if so desired,
by substituting the transformations associated with the preferred stress and strain measures in (\ref{eqn:transEU}--\ref{eqn:transBS2}).
\end{remark}

To complete the description of this multiscale material 
model, it is noted that for a static analysis or a dynamic analysis using an implicit time-stepping scheme, the consistent constitutive tangent of the plane stress constitutive law,
$\partial\widehat{\Sbm}_0^m/\partial\Ebm_0^m$, is typically required. This quantity is readily obtained using the constitutive tangent of the 3D constitutive law; its precise definition can be
found in \cite{klinkel2002using}.

\subsection{Discrete governing equations}
\label{subsec:fem}

Here, the discretized form of the equations governing the multiscale problem of interest are presented, notably including contact at both scales. Specifically,
\begin{itemize} [topsep=0pt] 
\itemsep-0.35em
\item At the macroscale, the solution of a dynamic contact problem is sought. The deforming bodies are discretized
      in space using membrane finite elements and in time using the explicit central difference time-integration
      scheme. The contact part of the problem is solved using an implicit approach \cite{salveson1996solution}.
\item At the microscale, the solution of static contact problems with prescribed displacement boundary conditions semi-discretized using solid finite elements is sought.
\end{itemize}
With regards to notation, a distinction is made in this work between \emph{unconstrained} degrees of
freedom (dofs), i.e., dofs that are not constrained by any essential boundary condition, and \emph{constrained}
dofs, i.e., dofs that are constrained by essential boundary conditions. A matrix or vector defined over
the set of unconstrained dofs is not designated by any specific symbol.  However, a vector of constrained
dofs is designated by the ring symbol as in $\mathring{\vbold}$ and a vector defined over the entire
set of constrained and unconstrained dofs is designated by the overline symbol as in $\bar\vbold$. In
other words,
\begin{equation}
\label{eqn:notation}
  {\bar\vbold} = \begin{bmatrix} \vbold \\ 
                                 \mathring{\vbold} \end{bmatrix}
\end{equation}

It is assumed, without loss of generality, that the discrete form of the governing macroscale equations can
be written as a differential-algebraic inequality (DAI) as follows
\begin{subequations}
\label{eqn:fem0}
\begin{align}
  \Mbm_0\ddot\ubm_0^{(n+1)} + \fbm_0^{int}\left(\bar\ubm_0^{(n+1)}\right) + \Gbm_0\left(\bar\ubm_0^{(n+2)}\right)\lambdabold_0^{(n+1)}
      &= \fbm^{ext}\left(t^{(n+1)}\right) \\ 
  \gbm_0\left(\bar\ubm_0^{(n+2)}\right) &\ge 0 \\
  \lambdabold_0^{(n+1)} &\le 0 \\
  {\lambdabold_0^{(n+1)}}^T \gbm_0\left(\bar\ubm_0^{(n+2)}\right) &= 0
\end{align}
\end{subequations}
where $\Mbm_0$ is the (diagonal) mass matrix, $\fbm_0^{int}$ and $\fbm_0^{ext}$ are the internal and
external force vectors, $\ubm_0^{(n+1)}$ and $\ddot\ubm_0^{(n+1)}$ are the displacements and accelerations
at time $t^{(n+1)}$, $\gbm_0$ is the gap, a vector-valued constraint function representing the discretized
non-penetration condition, $\Gbm_0$ is the transpose of the constraint Jacobian matrix
\begin{equation*}
  \Gbm_0 = \left[\frac{\partial\gbm_0}{\partial\ubm_0}\right]^T
\end{equation*}
and $\lambdabold_0^{(n+1)}$ is a vector of Lagrange multipliers at time $t^{(n+1)}$.
\begin{remark}
The evaluation of the proposed multiscale, plane stress constitutive function -- which encapsulates the microscale
response and its coupling with the macroscale counterpart -- is performed during the computation of the internal force vector
$\fbm_0^{int}$. Precisely, this computation is carried out in the same fashion as in the case of a conventional material law.
In particular, the contribution of each finite element to this computation is determined using an appropriate quadrature rule 
and the evaluation of the constitutive function at each quadrature point.
\end{remark}

Given some initial values $\ubm_0^{(n)}$, $\dot\ubm_0^{(n)}$ and $\ddot\ubm_0^{(n)}$ at time $t^{(n)}$,
the solution of the above inequality problem at time $t^{(n+1)}$ is obtained using the following updating procedure:
\begin{enumerate}[topsep=0pt]
\itemsep-0.35em
\item Update the displacement state
\begin{equation*}
  \ubm_0^{(n+1)} = \ubm_0^{(n)} + {\Delta t}_n \dot\ubm_0^{(n)} + 0.5 {\Delta t}_n^2 \ddot\ubm_0^{(n)}
\end{equation*}
\item Update the acceleration and velocity states using the predictor-corrector iterative method
\begin{enumerate}[topsep=0pt]
\itemsep-0.35em
\item predictor: $k = 0$
\begin{align*}
\ddot\ubm_0^{(n+1), \, 0} &= \Mbm_0^{-1}\left[\fbm_0^{ext}\left(t^{(n+1)}\right) - \fbm_0^{int}\left(\bar\ubm_0^{(n+1)}\right)\right] \\
\dot\ubm_0^{(n+1), \, 0} &= \dot\ubm_0^{(n)} + 0.5 {\Delta t}_n\left[\ddot\ubm_0^{(n)} + \ddot\ubm_0^{(n+1), \, 0}\right] \\
\ubm_0^{(n+2), \, 0} &= \ubm_0^{(n+1)} + {\Delta t}_{n+1} \dot\ubm_0^{(n+1), \, 0} + 0.5 {\Delta t}_{n+1}^2 \ddot\ubm_0^{(n+1), \, 0} \qquad
\end{align*}
\item corrector iterations: $k = 1, \ldots$
\begin{align*}
\ddot\ubm_0^{(n+1), \,k} &= \ddot\ubm_0^{(n+1), \,k-1} + \Delta \ddot\ubm_0^{(n+1), \,k} \\
\dot\ubm_0^{(n+1), \,k} &= \dot\ubm_0^{(n+1), \,k-1} + 0.5 \Delta t_n \Delta \ddot\ubm_0^{(n+1), \,k} \\
\ubm_0^{(n+2), \,k} &= \ubm_0^{(n+2), \,k-1} + 0.5 \left[ \Delta t_n \Delta t_{n+1} + \Delta t_{n+1}^2 \right] \Delta \ddot\ubm_0^{(n+1), \,k}
\end{align*}
\end{enumerate}
\end{enumerate}

At each corrector iteration, the acceleration increment $\Delta \ddot\ubm_0^{(n+1), \,k}$ is obtained by
linearizing the gap function $\gbm_0$ and solving the linearized subproblem
\begin{subequations}
\label{eqn:ICQP0}
\begin{align}
  \Mbm_0 \Delta \ddot\ubm_0^{(n+1), \,k} + \Gbm_0\left(\bar\ubm_0^{(n+2), \,k-1}\right) \lambdabold_0^{(n+1), \,k} &= -\tilde\fbm_0^{k-1} \\
  \Gbm_0\left(\bar\ubm_0^{(n+2), \,k-1}\right)^T \Delta \ddot\ubm_0^{(n+1), \,k} &\ge - \tilde\gbm_0^{k-1}  \\
  \lambdabold_0^{(n+1), \,k} &\le 0 \\
  {\lambdabold_0^{(n+1), \,k}}^T \; \left[ \Gbm_0\left(\bar\ubm_0^{(n+2), \,k-1}\right)^T \Delta \ddot\ubm_0^{(n+1), \,k} + \tilde\gbm_0^{k-1} \right] &= 0
\end{align}
\end{subequations}
where
\begin{align*}
  \tilde\fbm_0^{k-1} &= \Mbm_0 \left[ \ddot\ubm_0^{(n+1), \,k-1} - \ddot\ubm_0^{(n+1), \, 0} \right] \\
  \tilde\gbm_0^{k-1} &= \frac{2}{\Delta t_n \Delta t_{n+1} + \Delta t_{n+1}^2} \gbm_0\left(\bar\ubm_0^{(n+2), \,k-1}\right)
\end{align*}

The corrector subproblem (\ref{eqn:ICQP0}) has the form of a quadratic program: it can be solved by the
primal-dual active set method \cite{hintermuller2002primal, hueber2005primal}:
\begin{enumerate}[topsep=0pt]
\itemsep-0.35em
\item Initialize $\Delta\ddot\ubm_0^{(n+1), \,k}$, $\lambdabold_0^{(n+1), \,k}$
\item Iterate
\begin{itemize} [topsep=0pt] 
\itemsep-0.35em
\item Choose active set:
\begin{equation*}
  \mathcal{A} = \left\lbrace i : \left[ \lambdabold_0^{(n+1), \,k} \right]_i > 0 \, \wedge \,
                \left[\Gbm_0\left(\bar\ubm_0^{(n+2), \,k-1}\right)\right]_i^T \Delta\ddot\ubm_0^{(n+1), \,k} + \left[\tilde\gbm_0^{k-1}\right]_i < 0 \right\rbrace
\end{equation*}
\item Set the inactive Lagrange multipliers to zero:
\begin{equation*}
  \left[ \lambdabold_0^{(n+1), \,k} \right]_i = 0 \quad \forall \, i \, \notin \mathcal{A}
\end{equation*}
\item Solve for $\Delta\ddot\ubm_0^{(n+1), \,k}$ and the active Lagrange multipliers:
\begin{subequations}
\label{eqn:ECQP0}
\begin{align}
  \Mbm_0 \Delta \ddot\ubm_0^{(n+1), \,k} + \Gbm_0^\mathcal{A}\left(\ubm_0^{(n+2), \,k-1}\right)\lambdabold_0^{(n+1),\mathcal{A}, \,k} &= -\tilde\fbm_0^{k-1} \\
  \Gbm_0^\mathcal{A}\left(\ubm_0^{(n+2), \,k-1}\right)^T \Delta \ddot\ubm_0^{(n+1), \,k} &= - \tilde\gbm_0^{\mathcal{A}, \,k-1}
\end{align}
\end{subequations}
where the superscript $\mathcal{A}$ applied to a vector designates its restriction to the active set. Similarly,
the superscript $\mathcal{A}$ applied to a matrix designates its column-wise restriction to the active set.
\end{itemize}
\end{enumerate}

The active set method subproblem (\ref{eqn:ECQP0}) is a linear saddle-point system. To solve for the active
Lagrange multipliers, we first eliminate $\Delta \ddot\ubm_0^{(n+1), \,k}$ and then solve the remaining Schur
complement system
\begin{equation}
\label{eqn:sc1}
  \left[ {\Gbm_0^\mathcal{A}}^T \Mbm_0^{-1} \Gbm_0^\mathcal{A} \right] \lambdabold_0^\mathcal{A} 
  = \tilde\gbm_0^\mathcal{A} - {\Gbm_0^\mathcal{A}}^T \Mbm_0^{-1} \tilde\fbm_0.
\end{equation}
To simplify notation, the superscripts denoting the time-step index and predictor-corrector iteration have
been omitted here but can be inferred from (\ref{eqn:ECQP0}). After solving (\ref{eqn:sc1}) for the Lagrange
multipliers, the acceleration increment can be obtained from (\ref{eqn:ECQP0}a).

If $\Gbm_0^\mathcal{A}$ is rank-deficient, then the active set iterations may not converge. In this case,
a penalty parameter ($\mu$) can be used to regularize the system, leading to perturbed systems of
the form \cite{hintermuller2006path}
\begin{equation*}
  \left[ {\Gbm_0^\mathcal{A}}^T \Mbm_0^{-1} \Gbm_0^\mathcal{A} + \frac{1}{\mu} \Ibm \right] \lambdabold_0^\mathcal{A} 
  = \tilde\gbm_0^\mathcal{A} - {\Gbm_0^\mathcal{A}}^T \Mbm_0^{-1} \tilde\fbm_0
\end{equation*}
or equivalently,
\begin{equation*}
  \left[ \Mbm_0 + \mu\Gbm_0^\mathcal{A}{\Gbm_0^\mathcal{A}}^T \right] \Delta \ddot\ubm_0 
  = -\tilde\fbm_0 - \mu\Gbm_0^\mathcal{A}\tilde\gbm_0^\mathcal{A}.
\end{equation*}

This completes the description of the discrete macroscale problem and its solution algorithm.
Significantly, each time-step incurs only one evaluation of $\fbm^{int}$, which in the context of a multiscale
simulation invariably dominates the computational cost of the entire time-step. In order to evaluate this
discrete vector of internal forces, the homogenized stress tensor must be computed at each Gauss point
of the macroscale finite element model, which in turn involves the iterative solution of the Klinkel-Govindjee
plane stress equation with one solution of the discrete microscale governing equation required per iteration.
In the presence of contact at the microscale -- for example, non-penetration and sliding of yarns in a
woven fabric -- the discrete form of the microscale governing equation has a similar form to that of
the macroscale (\ref{eqn:fem0}) but without the time-dependence and associated temporal discretization.
The external force term is also identically zero and can be omitted; the problem is instead driven by
prescribed values of the constrained dofs and can be described as follows
\begin{subequations}
\label{eqn:fem1}
\begin{align}
  \fbm_1^{int}\left(\bar\ubm_1\right) + \Gbm_1\left(\bar\ubm_1\right)\lambdabold_1 &= 0 \\
  \gbm_1\left(\bar\ubm_1\right) &\ge 0 \\
  \lambdabold_1 &\le 0 \\
  \lambdabold_1^T \gbm_1\left(\bar\ubm_1\right) &= 0
\end{align}
\end{subequations}
All of the quantities $\Fbm_1$, $\bar\ubm_1$, $\gbm_1$, $\Gbm_1$ and $\lambdabold_1$ are microscale counterparts
of the corresponding macroscale quantities previously defined. The above problem can be solved in a similar fashion
to that of the macroscale problem by solving a series of linearized subproblems of the form
\begin{subequations}
\label{eqn:ICQP1}
\begin{align}
  \Kbm_1^{tgt} \Delta \ubm_1^{k} + \Gbm_1\left(\bar\ubm_1^{k-1}\right) \lambdabold_1^{k} &= -\tilde\fbm_1^{k-1} \\
  \Gbm_1\left(\bar\ubm_1^{k-1}\right)^T \Delta \ubm_1^{k} &\ge - \tilde\gbm_1^{k-1}  \\
  \lambdabold_1^{k} &\le 0 \\
  {\lambdabold_1^{k}}^T \; \left[ \Gbm_1\left(\bar\ubm_1^{k-1}\right)^T \Delta \ubm_1^{k} + \tilde\gbm_1^{k-1} \right] &= 0
\end{align}
\end{subequations}
where $\Kbm_1^{tgt}$ is the microscale tangent stiffness matrix
\begin{equation*}
  \Kbm_1^{tgt} = \frac{\partial\fbm_1^{int}}{\partial\ubm_1}
\end{equation*}
Problem (\ref{eqn:ICQP1}) can again be solved by the dual-primal active set method proposed for the corresponding
macroscale problem (\ref{eqn:ICQP0}), although numerous alternatives exist.

The computational homogenization method described herein provides a very general
framework for solving the problem of interest without resorting to any ad-hoc approximation. However,
without introducing any further approximation, the framework -- although amenable to parallel implementation --
is impractical for all but the most modest of applications due to its computational complexity. For example,
the authors of this paper estimate that to simulate the inflation of a parachute using a macroscale model comprising 182,554 nodes
and 279,025 triangular membrane elements would require 49,604,444,444 constitutive function evaluations
and a total run time of approximately 48 years on 1,000 processing units. Hence, a regression-based surrogate modeling methodology is proposed
to achieve computational tractability and described below. It is emphasized that this methodology relies exclusively on
the general framework presented above to obtain ``training data'' that can be used to construct a low-dimensional surrogate model.

\section{Regression-based surrogate microscale model}
\label{sec:micro-surrogate}

Here, a methodology featuring a regression-based surrogate model is presented for dramatically accelerating the solution of nonlinear dynamic multiscale problems modeled using the multiscale formulation 
based on the concept of a locally attached microstructure overviewed above. The methodology features a novel training strategy based on the concept of a coupon test analogy. 

Regression-based surrogate models can be loosely classified as follows:
\begin{enumerate}[topsep=0pt]
\itemsep-0.35em
\item Models whose forms are determined {\it a priori} and whose parameters are fitted to available data. Examples of such models are:
\begin{itemize}[topsep=0pt]
\itemsep-0.35em
\item The St. Venant-Kirchhoff hyperelastic model, a two-parameter model characterized by a linear relationship between the second Piola-Kirchhoff stress and the Green-Lagrange strain.
\item Hyper-viscoelastic models incorporating a hyperelastic model such as the St. Venant-Kirchhoff model, combined with a viscoelastic component based on a Prony series.
\end{itemize}
\item Models whose forms are not entirely predetermined but which are rather discovered, at least in part, by a regression/fitting process. An example of such a model is an artificial 
NN-based model. In this case, certain characteristics of the model may still be specified {\it a priori}, such as the number of hidden layers and the functional form of the 
activation function. 
\end{enumerate}

A training strategy, i.e., a procedure for sampling a parameter space such as $\left \{ E_0^{(11)}, \, E_0^{(12)}, \, E_0^{(22)} \right \}$ and collecting conjugate stress and strain data for the 
purpose of constructing a regression-based constitutive model is proposed here.
The strategy employs a small coupon of the material of interest that is semi-discretized at the macroscale level by a single membrane element. It is emphasized that due to the overwhelming cost 
of an entire multiscale simulation based on a high-dimensional macroscale model, it is not practical to collect data specifically customized to a target application, as is sometimes done to train 
projection-based reduced-order models (PROMs). However, to some extent, the range of strains to which the coupon model is subjected to during the training can be customized, for example, to target 
applications with small, medium, or large deformations. Due to the small size of the coupon macroscale model, it is feasible to collect data comprehensively sampled on a regular grid within 
a three-dimensional parameter space such as  $\left \{ E_0^{(11)}, \, E_0^{(12)}, \, E_0^{(22)} \right \}$ (recall that the prescribed microscale boundary displacements are obtained from a mapping from 
the in-plane components of the macroscale symmetric Green-Lagrange strain tensor). For training purposes, the macroscale strain can be indirectly specified by prescribing displacements on the 
boundary of the macroscale model of the coupon. More importantly, the generation and collection of multiscale data can be accelerated using the nonlinear PMOR approach described in
Appendix \ref{sec:micro-rom}.

Figure \ref{fig:micro-snaps} shows for several points sampled in the parameter space $\left \{ E_0^{(11)}, \, E_0^{(12)}, \, E_0^{(22)} \right \}$,
deformed configurations and corresponding von Mises stress contours obtained during a training performed for the application described in the following section.

\begin{figure}[ht]
\centering{
\includegraphics[width=0.48\linewidth]{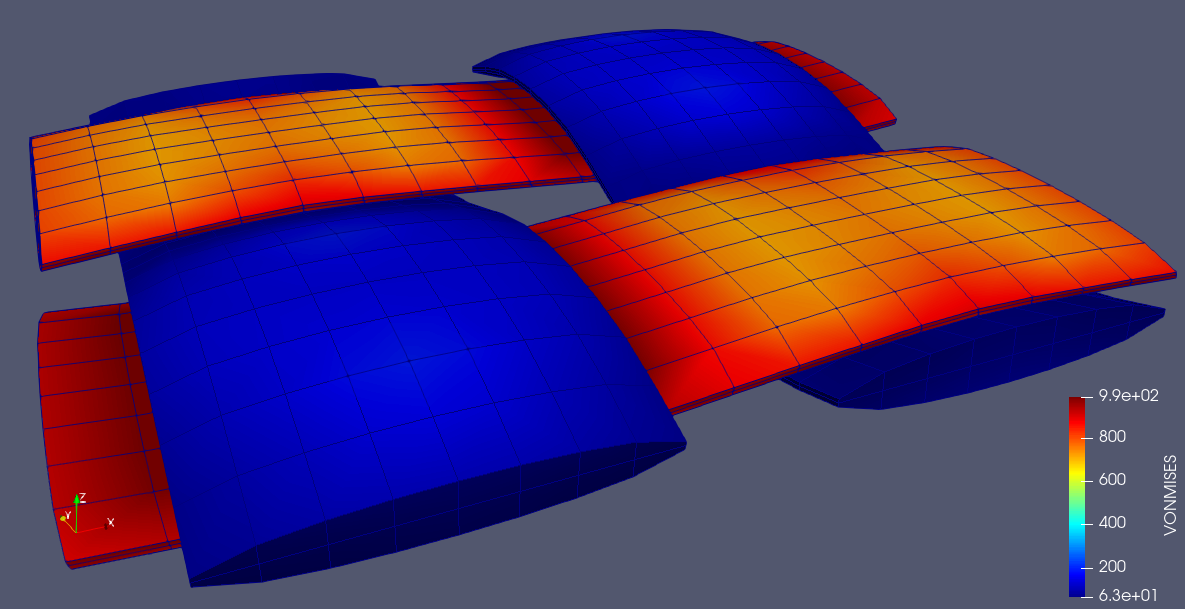}~~
\includegraphics[width=0.48\linewidth]{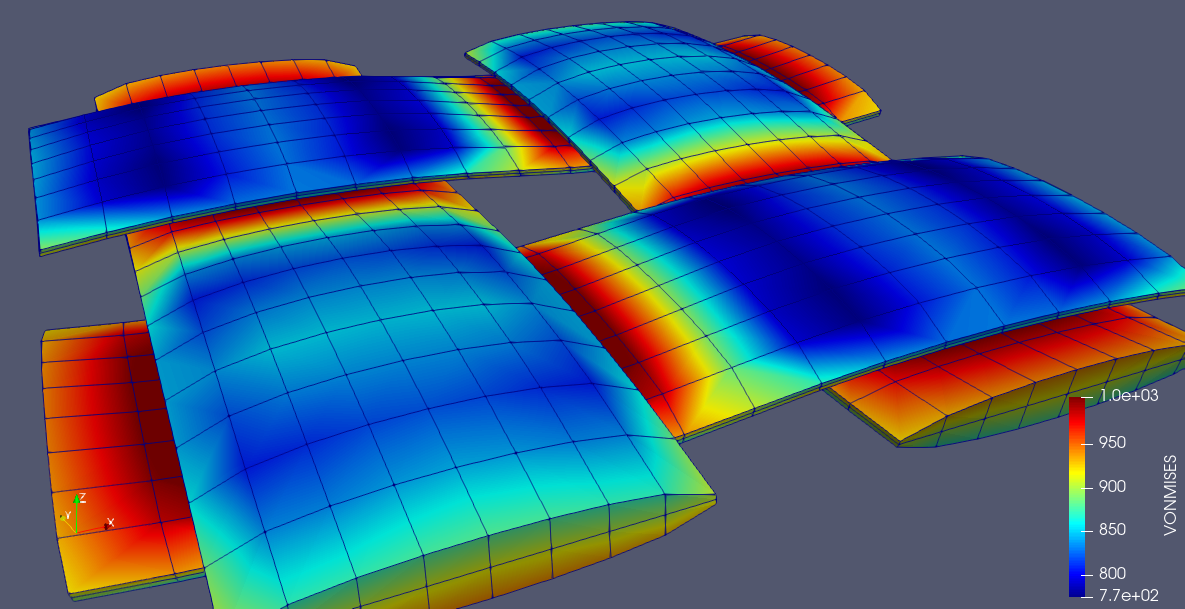}\\
\vspace{8pt}
\includegraphics[width=0.48\linewidth]{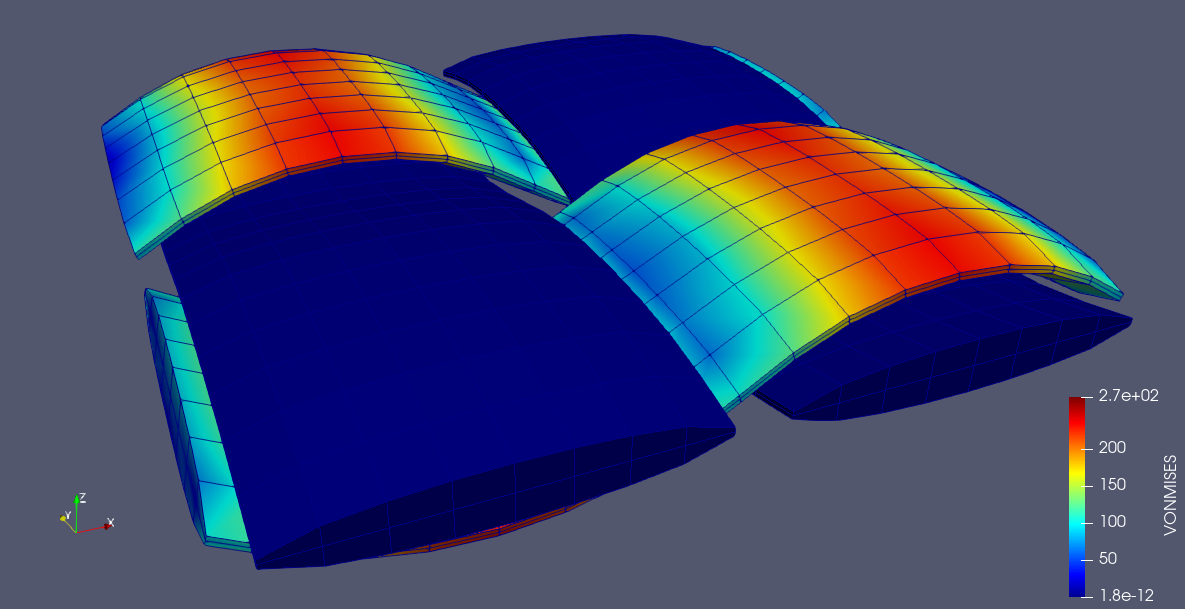}~~
\includegraphics[width=0.48\linewidth]{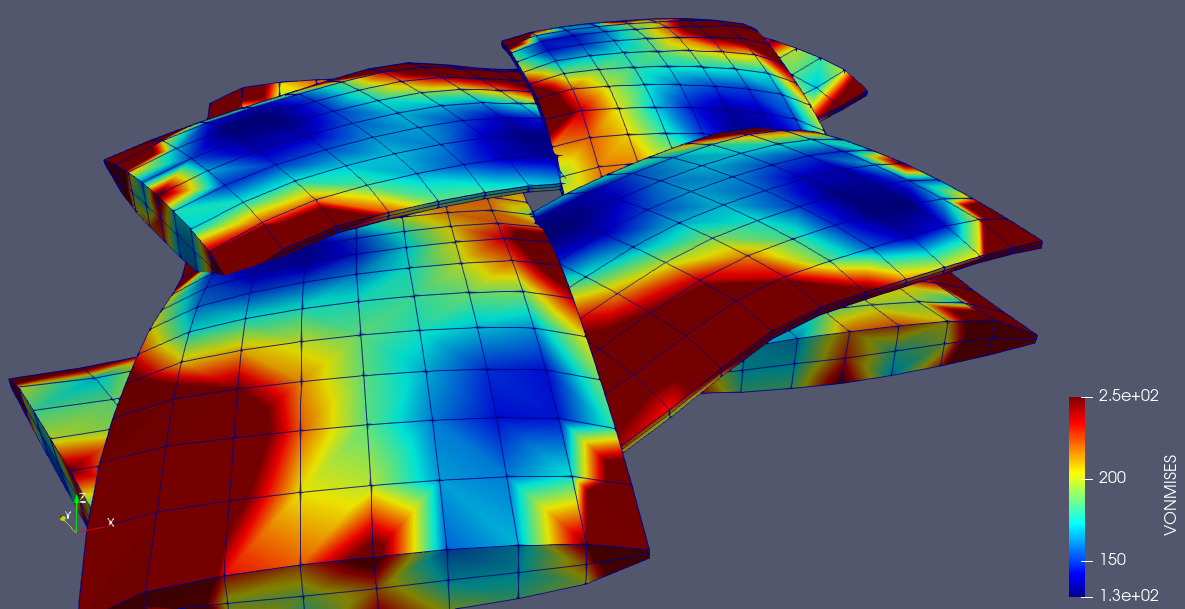}
} 
\caption{Microscale training displacement solution snapshots colored by von Mises stress contours for a few points sampled in the parameter space 
$\left \{ E_0^{(11)}, \, E_0^{(12)}, \, E_0^{(22)} \right \}$: uniaxial tension $\left\{0.16, 0., 0.\right\}$ (top left); biaxial tension
	$\left\{0.15, 0.15, 0.\right\}$ (top right); uniaxial compression $\left\{-0.06, 0., 0.\right\}$ (bottom left); and shear $\left\{0., 0., 0.125\right\}$ (bottom right).}
\label{fig:micro-snaps}
\end{figure}

\section{Applications}
\label{sec:app}

In this section, the computational homogenization framework proposed in this paper for the nonlinear dynamic analysis of membranes is demonstrated for the coupled, multiscale,
fluid-structure simulation of the supersonic inflation of a DGB parachute canopy made of a woven fabric during a Mars landing event. For this purpose, three regression-based surrogate models are 
considered to accelerate the microscale computations:
\begin{itemize}[topsep=0pt] 
\itemsep-0.35em
\item A linear regression model defined by
\begin{equation}
\label{eq:linear}
  \begin{bmatrix} S_0^{(11)} & S_0^{(22)} & S_0^{(12)} \end{bmatrix}^T = \boldsymbol{C}_l \begin{bmatrix} E_0^{(11)}
                  & E_0^{(22)} & 2\hspace{-.1em}E_0^{(12)} \end{bmatrix}^T
\end{equation}
and the symmetric matrix $\boldsymbol{C}_l {\in \mathbb R}^{3\time 3}$.
\item A quadratic regression model defined by
\begin{equation}
\label{eq:quadratic}
\resizebox{.9\hsize}{!}{$
  \begin{bmatrix} S_0^{(11)} & S_0^{(22)} & S_0^{(12)} \end{bmatrix}^T = \boldsymbol{C}_q 
  \begin{bmatrix} E_0^{(11)} &\; E_0^{(22)} &\; 2\hspace{-.1em}E_0^{(12)} &\; {E_0^{(11)}}^2 &\; {E_0^{(22)}}^2 &\;
                  \left(2\hspace{-.1em}E_0^{(12)}\right)^2 &\; 2\hspace{-.1em}E_0^{(12)}\hspace{-.25em}E_0^{(22)} &\;
                  2\hspace{-.1em}E_0^{(12)}\hspace{-.25em}E_0^{(22)} &\; E_0^{(11)}\hspace{-.25em}E_0^{(22)}
  \end{bmatrix}^T
$}
\end{equation}
and $\boldsymbol{C}_q {\in \mathbb R}^{3\times 9}$.
\item A linear model with a NN-based correction of the form
\begin{equation}
\label{eq:NN}
  \begin{bmatrix} S_0^{(11)} & S_0^{(22)} & S_0^{(12)} \end{bmatrix}^T = \boldsymbol{C}_l \begin{bmatrix} E_0^{(11)}
                  & E_0^{(22)} & 2\hspace{-.1em}E_0^{(12)} \end{bmatrix}^T + \boldsymbol{N}\left( \begin{bmatrix} E_0^{(11)}
                  & E_0^{(22)} & 2\hspace{-.1em}E_0^{(12)} \end{bmatrix}^T\right)
\end{equation}
\end{itemize}
where $\boldsymbol{N}$ denotes a fully connected NN designed for correcting the linear model by mapping the strain to a stress correction. 

First, it will be shown that the NN-based model outlined above outperforms the two other surrogate models in terms of training and test errors. For this reason, only this model will be considered
in Section \ref{sec:parachute} to describe the behavior of the canopy material in the simulation of the supersonic parachute inflation of the DGB parachute.

\subsection{Artificial neural networks}
\subsubsection{Data generation}
\begin{figure}[ht]
\centering
\begin{tikzpicture}[scale=0.4]
\draw (0,0) -- (7.5,0) -- (0, 7.5) -- (0, 0);
\draw (-0.3, - 0.51961524227) -- (0, 0) -- (0.3, -0.51961524227) -- (-0.3, - 0.51961524227);
\draw[thick,->] (7.5, 0) -- (8.5, 1.0) node[anchor=south]{$\mathring{\bm{u}}_0$};
\draw[thick,->] (0, 7.5) -- (1.0, 8.5) node[anchor=south]{$\mathring{\bm{u}}_0$};
\end{tikzpicture}
\caption{Schematic of the numerical coupon test with prescribed displacements (strain fields).}
\label{fig:coupon_test}
\end{figure}
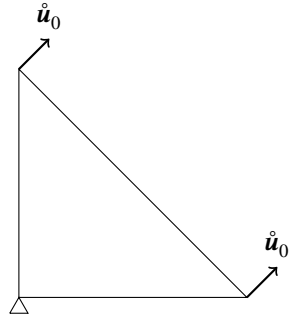

Different stress-strain tensor data pairs $\left(\boldsymbol{S}_{0}^{(i)},\boldsymbol{E}_{0}^{(i)}\right)$, $i=1,\ldots,N$ are generated by performing a numerical coupon test $N$ times, where $N$ is 
the number of training data points. Each coupon test is graphically depicted in Figure \ref{fig:coupon_test}, where the right triangle geometry representing a single finite element
has two sides of length equal to $1$~m. The displacements of the right angle node and all out-of-plane displacements are constrained to be zero; prescribed in-plane displacements are applied to the 
two other nodes to generate a specified target strain field. 

Each microscale problem (see Section \ref{subsec:sb}) is solved at the single Gauss quadrature point located at the center of the right triangle using the nonlinear PMOR approach
described in Appendix \ref{sec:micro-rom}, which accelerates the generation of the homogenized strain and stress pairs. 

\begin{figure}[ht]
\centering{
\includegraphics[width=0.6\linewidth]{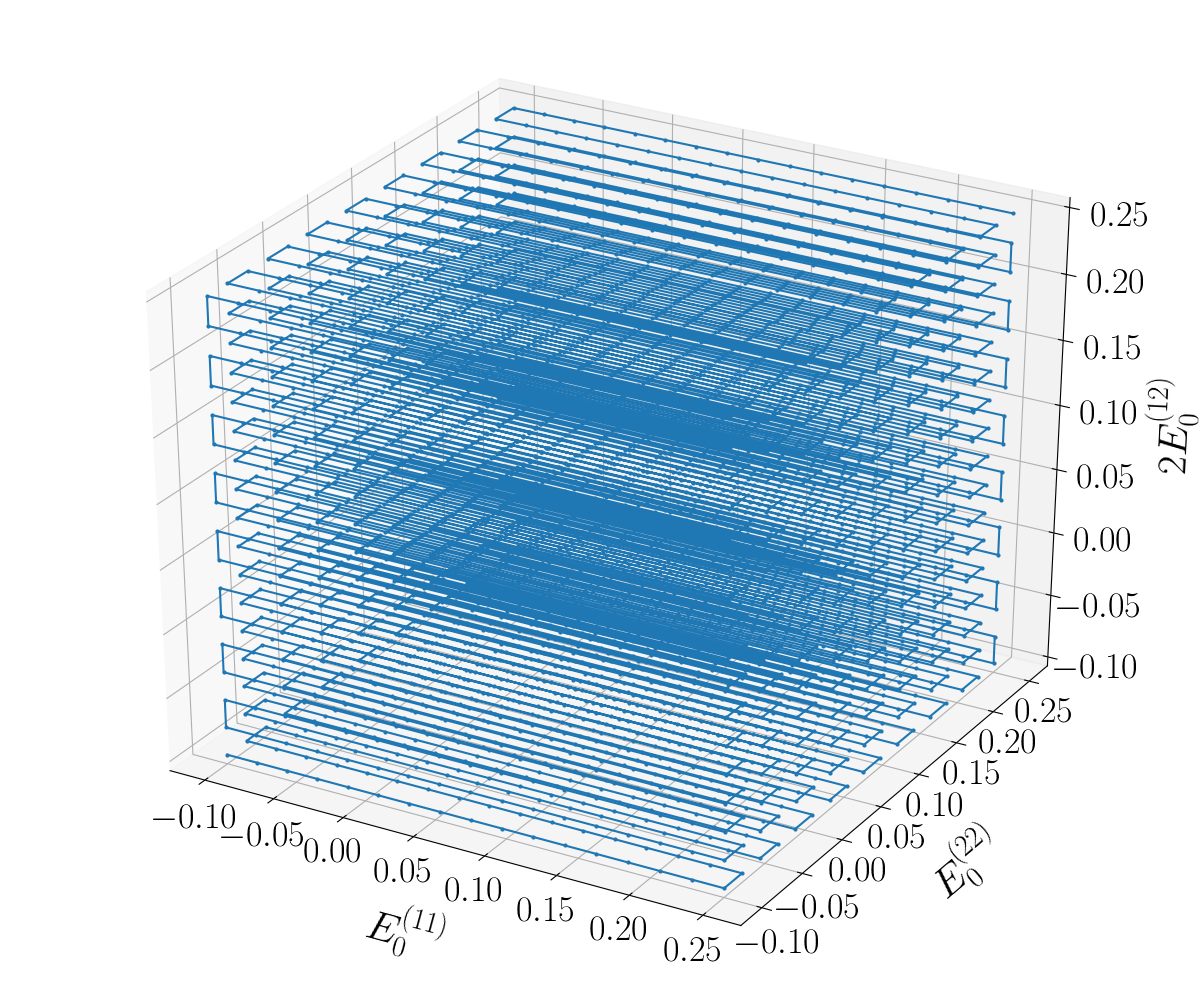}
}
\caption{Trajectory used for sampling training data points.}
\label{fig:trajectory}
\end{figure}

The strain field at the macroscale (fabric coupon) level $\begin{bmatrix} E_0^{(11)} & E_0^{(22)} & 2E_0^{(12)}
\end{bmatrix}^T$ is sampled in a cube of extent $[-0.1, 0.25] \times [-0.1, 0.25] \times [-0.1, 0.25]$.
Here, the range of strains is customized to match the application of interest, specifically, the supersonic inflation
of a DGB parachute discussed in Section \ref{sec:parachute}. The cube is uniformly sampled using $17$ equidistant
points in each strain component, which accounts for a total of $4,913$ training data points. Each training data point
requires the solution of the discrete equations (\ref{eqn:fem0}) governing the multiscale coupon problem.
To facilitate the implementation of the sampling procedure, each data point is generated as a time-step of a single multiscale
simulation in which the prescribed boundary conditions are varied in time along the trajectory shown in Figure
\ref{fig:trajectory}. Each time-step can be interpreted as an independent static simulation; alternatively, each line segment of 
the trajectory can be interpreted as being associated with the numerical counterpart of a single physical coupon test in which two 
strain components are held fixed, while the third is varied. Crucially, the converged solution of the microstructure problem
at the previous data point is used to initialize Newton's method at the next data point. In total, the multiscale data generation procedure equipped with the nonlinear
PMOR approach described in Appendix \ref{sec:micro-rom}
consumes about 40 hours wall-clock time on a single core. To validate the surrogate models, another set of $4,913$ test data points is also generated by shifting the 
aforementioned trajectory.

\subsubsection{Training}
\label{sec:sst}
For the sake of computational efficiency, the number of hidden layers for the NN introduced in (\ref{eq:NN}) is set to 1.
Both tanh and ReLU activation functions are considered. 

\begin{figure}[ht]
\centering{
\includegraphics[width=0.75\linewidth]{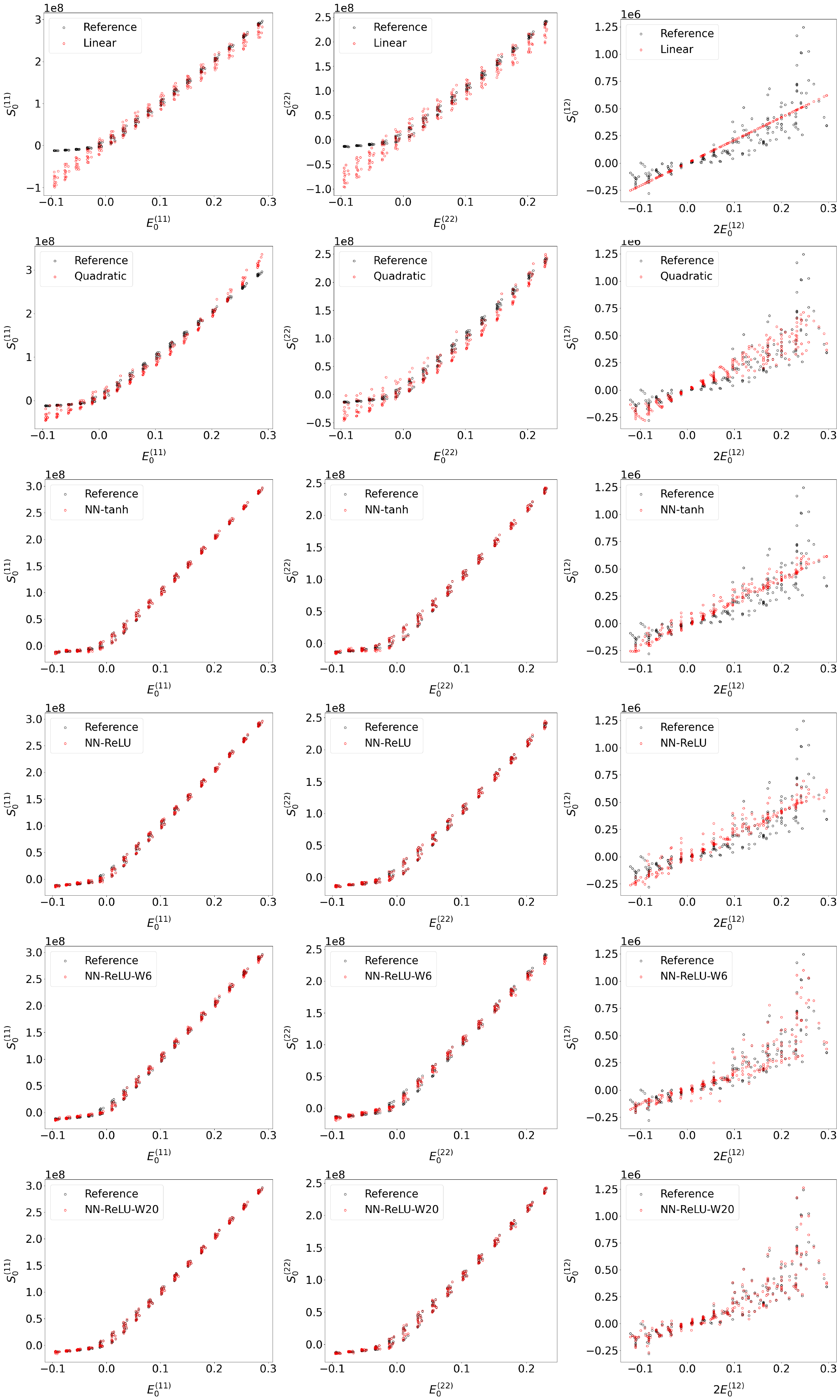}
}
\caption{Stress-strain curves predicted using the exact (reference) and surrogate microscale models (training data set).}
\label{fig:Train-E-S}
\end{figure}

A good definition of the loss function can be
\begin{equation}
  \sum_{i=1}^{N}||\bm{S}_0^{(i)} - \mathcal{M}(\bm{E}_0^{(i)})||_2^2 + \lambda ||\bm{\theta}||_2^2
\label{eq:loss1}
\end{equation}
where $\mathcal{M}$ represents the surrogate model, $\bm{\theta}$ denotes its hyperparameters and $L_2$ regularization is added using the parameter $\lambda$, which is set here to $\lambda = 10^{-4}$. 
For a woven fabric material however, the shear stress $S_0^{(12)}$ is typically several orders of magnitude smaller than the axial stresses $S_0^{(11)}$ and $S_0^{(22)}$ 
(see Figure \ref{fig:Train-E-S}). For this reason, the alternative weighted loss function 
\begin{equation}
  \sum_{i=1}^{N}\left(S_0^{(11),(i)} - \mathcal{M}^{(11)}(\bm{E}_0^{(i)})\right)^2 
  + \left(S_0^{(22),(i)} - \mathcal{M}^{(22)}(\bm{E}_0^{(i)})\right)^2 + w^2\left(S_0^{(12),(i)}
  - \mathcal{M}^{(12)}(\bm{E}_0^{(i)})\right)^2 + \lambda ||\bm{\theta}||^2
\label{eq:loss2}
\end{equation}
where $w$ is a weighting constant is more appropriate. For a given training data set, the value of $w$ can be automatically deduced from the application to this set of a simple scaling procedure. 

In total, six regression-based surrogate models are considered: 
\begin{itemize} [topsep=0pt] 
\itemsep-0.35em
  \item The linear model (\ref{eq:linear}).
  \item The quadratic model (\ref{eq:quadratic}).
  \item The NN-tanh model, which is model (\ref{eq:NN}) where the NN is equipped with six tanh neurons and trained with the non-weighted loss function (\ref{eq:loss1}) (for comparison only).
  \item The NN-ReLU model, which is model (\ref{eq:NN}) where the NN is equipped with six ReLU neurons and trained with the non-weighted loss function (\ref{eq:loss1}) (for comparison only).
  \item The NN-ReLU-W6 model, which is model (\ref{eq:NN}) where the NN is equipped with six ReLU neurons and trained with the weighted loss function (\ref{eq:loss2}).
  \item The NN-ReLU-W20 model, which is model (\ref{eq:NN}) where the NN is equipped with 20 ReLU neurons and trained with the weighted loss function (\ref{eq:loss2}).
\end{itemize} 
Both linear and quadratic models are trained without regularization ($\lambda = 0$). All NNs are trained using the limited-memory BFGS (L-BFGS-B) method \cite{byrd1995limited} with 
regularization ($\lambda = 10^{-4}$). In all cases, the line search routine provided in \cite{more1994line} is used: it attempts to enforce the Wolfe conditions \cite{byrd1995limited} using a sequence 
of polynomial interpolations. Note that the BFGS algorithm is appropriate in this case because the data sets are relatively small \cite{huang2020predictive}; for larger data sets,
the stochastic gradient descent method is suggested for training.

\begin{figure}[ht]
\centering{
\includegraphics[width=0.75\linewidth]{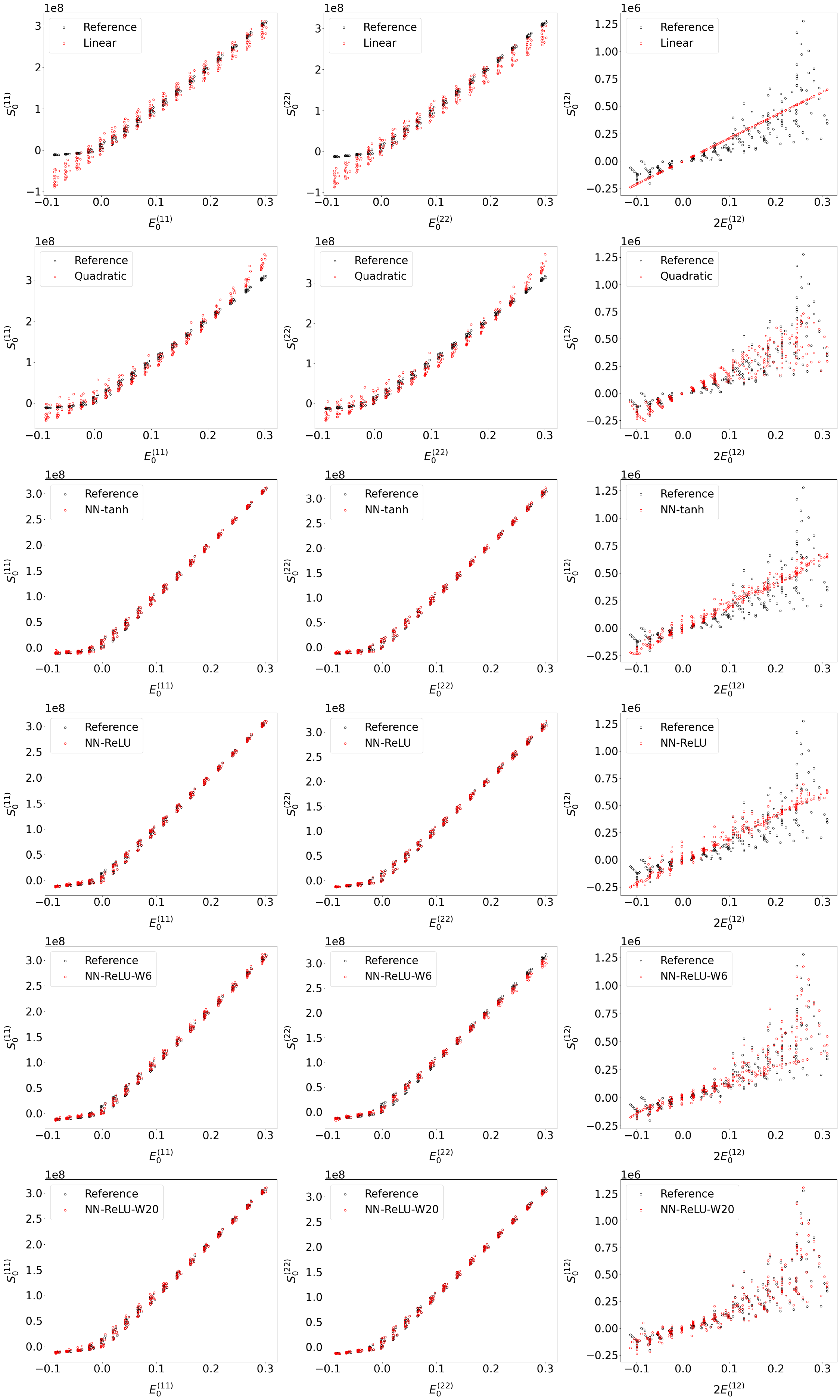}
}
\caption{Stress-strain curves predicted using the exact (reference) and surrogate microscale models (test data set).}
\label{fig:Test-E-S}
\end{figure}

For both training and test data sets, the relative total errors and relative errors for each stress component are reported in Table \ref{tab:errors}. The reader can observe 
that the best if not all NN-based surrogate models lead 
to relative errors that are one order of magnitude smaller than those of the linear and quadratic regression models. The training/test data and all obtained
predictions are also plotted in Figure \ref{fig:Train-E-S} and Figure \ref{fig:Test-E-S}, for each of the component-wise relations $S_0^{(11)}-E_0^{(11)}$, $S_0^{(22)}-E_0^{(22)}$, and 
$S_0^{(12)}-2E_0^{(12)}$. The data shows that the woven fabric material is flexible with respect to shear and compression. In particular, the shear stresses are found to be two orders of magnitude 
smaller than the axial stresses under similar strains. Furthermore, the $S_0^{(11)}-E_0^{(11)}$ and $S_0^{(22)}-E_0^{(22)}$ curves are ``flat'' when the woven fabric is compressed, indicating that it
does not take compression; their slopes suddenly change at 
zero and remain constant in the stretching regime. Due to these features (especially slope discontinuity in the stress-strain relations), the NN-based regression models deliver better approximations 
and therefore outperform the linear and quadratic regression models. 

It is worth mentioning that the shear stress is relatively small but highly nonlinear. NNs trained with the non-weighted loss function focus mainly on the axial stresses and therefore fail to capture 
the nonlinearity in the shear stress (see Figure \ref{fig:Train-E-S} and Figure \ref{fig:Test-E-S}). On the other hand, the NNs trained with the weighted loss function deliver a reasonable accuracy 
for shear stress prediction. Moreover, increasing the number of neurons is shown to improve accuracy (see Table \ref{tab:errors}).

\begin{table}[ht]
\centering
\begin{tabular}{c|c|c|c|c|c|c}
\toprule
             & Linear & Quadratic & NN-tanh & NN-ReLU & NN-ReLU-W6 & NN-ReLU-W20\\
\midrule
Training set & 19.5\% & 10.4\%    & 1.06\%  & 1.05\%  & 2.13\%  & 0.97\% \\
$S_{xx}$     & 19.5\% & 10.4\%    & 0.93\%  & 1.10\%  & 1.63\%  & 0.88\% \\
$S_{yy}$     & 19.5\% & 10.4\%    & 1.24\%  & 0.97\%  & 2.74\%  & 1.08\% \\
$S_{xy}$     & 42.4\% & 38.6\%    & 39.1\%  & 38.3\%  & 19.2\%  & 8.20\% \\
\midrule
Test set     & 14.0\% & 9.87\%    & 0.91\%  & 1.03\%  & 2.40\%  & 0.95\% \\
$S_{xx}$     & 14.0\% & 9.87\%    & 0.91\%  & 1.17\%  & 1.83\%  & 1.00\% \\
$S_{yy}$     & 14.0\% & 9.87\%    & 0.91\%  & 0.87\%  & 2.83\%  & 0.90\% \\
$S_{xy}$     & 43.0\% & 34.81\%   & 38.7\%  & 39.2\%  & 21.6\%  & 9.37\% \\
\bottomrule
\end{tabular}
\caption{Relative total errors and relative errors for each stress component and each considered regression-based surrogate model, for both training and test data sets.}
\label{tab:errors}
\end{table}

Regarding computational cost, the number of operations performed during a single evaluation of a surrogate model
is $O(15)$ for the linear regression model, $O(51)$ for the quadratic one and $O(13\times \#\textrm{neurons} + 15)$ for the NN-based regression
models. For an explicit time-integration, computational cost is typically dominated by that associated with constitutive function evaluations.
Hence, in the context an explicit time-integrator, adopting an NN-based surrogate microscale model instead of a linear one may increase the cost of
a multiscale simulation by up to a factor of 6 in the case of 6 neurons, or 18 in the case of 20 neurons. However, in the presence of
an implicit time-integration scheme, the computational cost associated with an equation solver is typically such that the additional cost incurred by an NN-based 
regression model over a linear one would be substantially less. In any case, the realistic, coupled, multiscale, fluid-structure simulation discussed in the next 
section is not computationally tractable without a surrogate microscale model. For this reason, and because of its superior accuracy, the surrogate model
NN-ReLU-W20 is chosen for performing this simulation -- and model NN-ReLU is considered only for the purpose of performing a comparison.

\clearpage

\subsection{Supersonic inflation of a disk-gap-band parachute for Mars landing}
\label{sec:parachute}
Finally, the  proposed computational homogenization framework is equipped with the previously trained NN-ReLU models and applied here to simulate the supersonic
inflation dynamics of a NASA DGB parachute system in the low-density, low-pressure, supersonic Martian atmosphere \cite{cruz2014reconstruction, huang2020embedded, huang2020modeling}.
While such a coupled, multiscale, fluid-structure simulation is crucial to the understanding of the effects of a woven fabric material on the performance of a parachute during the deceleration process, 
its main purpose here is two-fold: 1) demonstrate the computational tractability of the proposed computational framework for a realistic application; and 2) to \emph{validate} (partially) it
using flight data from the landing on Mars of NASA's rover Curiosity.

\begin{figure}[ht]
\centering
\includegraphics[width=1.0\textwidth]{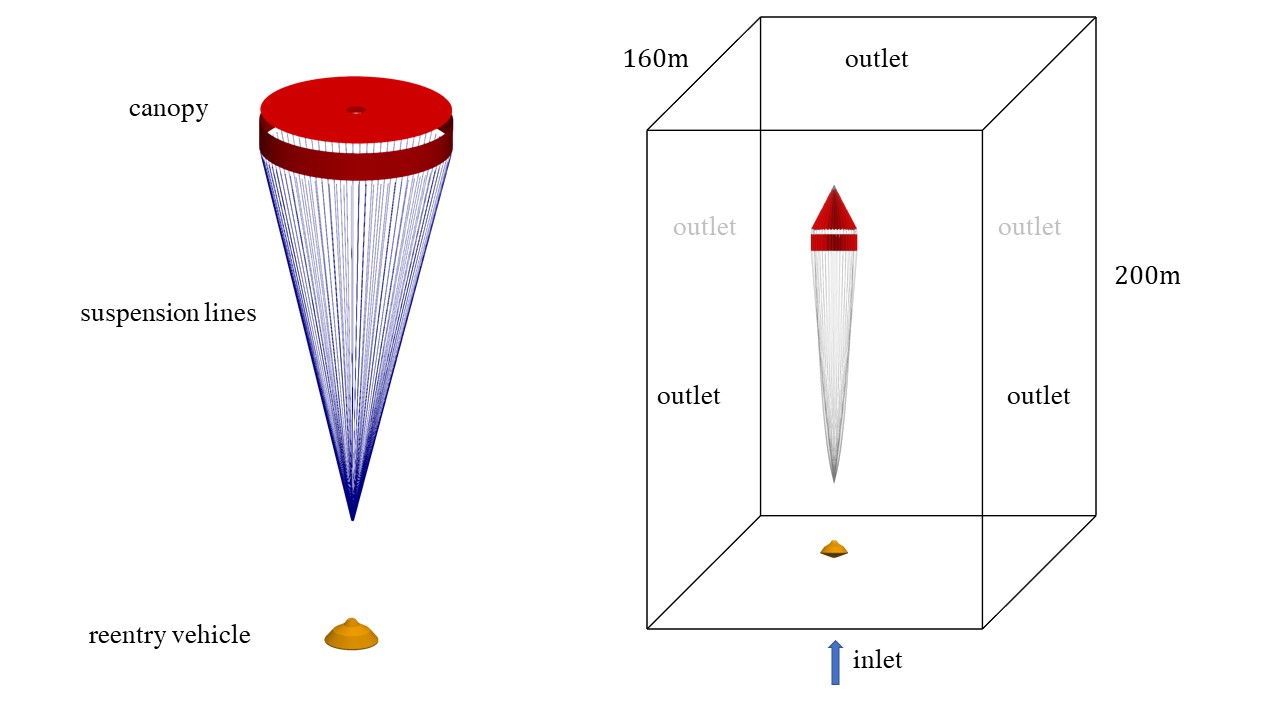}
\caption{Simulation of the supersonic parachute inflation dynamics of a NASA DGB system: system configuration (left); and embedding computational fluid domain and embedded initial folded configuration 
(right).}
\label{fig:PIDCFD}
\end{figure}

Specifically, the DGB parachute system considered here is that which successfully deployed in 2012 for the Mars landing of Curiosity (see Figure \ref{fig:PIDCFD}-left). This aerodynamic decelerator 
system consists of three main components \cite{cruz2014reconstruction}:
\begin{itemize}[topsep=0pt] 
\itemsep-0.35em
\item The canopy, which is made of a woven nylon fabric material (see Figure \ref{fig:figure1}).
\item The suspension lines, which are made of Technora T221 braided cords.
\item The reentry vehicle.
\end{itemize}
Its geometric and material properties are listed in Table \ref{tab:parachute}.  

\begin{table}
\centering
\begin{tabular}{@{}llll@{}}
\toprule
Component &  Parameter & Description & Value\\ 
\midrule
Canopy & $D$      & diameter        & 15.447 m                     \\
       & $t$      & thickness       & 7.6 $\times$ 10$^{-5}$ m     \\
       & $E$      & \emph{microscale} yarn Young's modulus &  3497 MPa
       \tablefootnote{The microscale yarn Young's modulus is roughly estimated from the Young's modulus of the macroscale nylon.} \\
       & $\nu$    & \emph{microscale} yarn Poisson's ratio & 0.2   \\
       & $\rho^C$ & density         & 1154.25 kg m$^{-3}$          \\
       & $\alpha$ & porosity        & 0.08                         \\ 
\midrule
Suspension lines & $L$         & length          & 36.56 m                    \\
                 & $D$         & diameter        & 3.175 $\times$ 10$^{-3}$ m \\
                 & $E$         & Young's modulus & 29.5 GPa                   \\
		 & $\rho^{SL}$ & density         & 1154.25 kg m$^{-3}$        \\
\bottomrule
\end{tabular}
\caption{Simulation of the supersonic parachute inflation dynamics of a NASA DGB system: geometric and material properties of the system \cite{lin2010flexible,cruz2014reconstruction, hill2016mechanical}.}
\label{tab:parachute}
\end{table}

The simulation discussed herein starts from the line stretch stage where the suspension line subsystem is deployed, the canopy is folded (see Figure \ref{fig:PIDCFD}-right)
and the entire system is prestressed by the folding pattern \cite{huang2020modeling}. The incoming supersonic flow is at the state defined by
the free-stream conditions $M_{\infty} = 1.8$, $\rho_{\infty} = 0.0067$~kg~m$^{-3}$ and $p_{\infty} = 260$~Pa.

Since the Martian atmosphere is mainly composed of carbon dioxide, the viscosity of this gas is modeled
using Sutherland's viscosity law with the constant $\mu_0 = 1.57\times10^{-6}$~kg m$^{-1}$s$^{-1}$ and
the reference temperature $T_0 = 240$~K. The Reynolds number based on the canopy diameter is $4.06 \times 10^6$.
Hence, at the beginning of the simulation, the flow is assumed to have transitioned to the turbulent regime -- which is modeled here using
Vreman's eddy viscosity subgrid-scale model for turbulent shear flow \cite{vreman2004eddy} equipped with the model constant $C_s = 0.07$. 

Given the expected large motions and deformations of the parachute system during its inflation, the flow computations are performed
using the large eddy simulation (LES) capability of the AERO-F flow solver \cite{geuzaine2003aeroelastic, farhat2003application} and its embedded boundary method for fluid-structure interaction
known as the finite volume method with exact two-material Riemann problems (FIVER) \cite{wang2011algorithms, farhat2012fiver, lakshminarayan2014embedded, wang2015computational,
main2017enhanced, huang2020embedded}. AERO-F incorporates a parallel Adaptive Mesh Refinement (AMR) capability based on newest vertex bisection \cite{mitchell1988unified, borker2019mesh}, which enables 
it to capture various interactions between the fluid system, the nonlinear parachute system including its suspension lines and the forebody. 

The canopy of the DGB parachute consists of band and disk gores that are represented here by 279,025 geometrically nonlinear membrane elements. The suspension line subsystem contains 80 lines, 
each of which is discretized by 500 geometrically nonlinear beam elements. The reentry vehicle is modeled as a fixed rigid body: it is embedded, together with the entire aerodynamic decelerator system, 
in the embedding computational fluid domain (see Figure \ref{fig:PIDCFD}). This domain is a box of size 200 m $\times$ 160 m $\times$ 160 m. It is initially discretized by a mesh with
2,778,867 nodes and 16,308,672 tetrahedra. During the fluid-structure-interaction simulation, AMR is applied to track and resolve the boundary layers and flow features. For this purpose,
the specified characteristic mesh sizes near the reentry vehicle and canopy are $2.5$~cm and $5$~cm, respectively: those in the wake and near the shock are set to $10$~cm.  

Since the canopy is made of a woven nylon fabric with an $8\%$ void fraction, its permeability is modeled using a homogenized porous wall model \cite{huang2019homogenized, huang2018simulation}. Due to 
the massive self-contact of the parachute canopy during its dynamic inflation, the explicit central difference time-integration scheme is used to advance in time the semi-discrete state of the structural
system. A small amount of Rayleigh damping is applied to stabilize this system.

First, a quasi-steady state of the flow past the folded parachute configuration shown in Figure \ref{fig:PIDCFD}-right
is computed assuming that this configuration is rigid and fixed. Using the computed CFD solution and the aforementioned
prestressed state of the structural model of the parachute system as initial fluid and structural conditions,
respectively, the coupled, multiscale, fluid-structure simulation of the inflation dynamics of the DGB parachute is performed in the time-interval 
$[0, 0.8]$~s. The length of this time-interval is such that it covers the inflation process as well as
a few breathing cycles of the DGB parachute system. While the explicit central difference time-integrator is applied to advancing in time the semi-discrete structural system
for the reason mentioned above, the implicit, 3-point backward difference formula (BDF) scheme is applied to time-integrate the semi-discrete fluid state. 
The fluid and structural discretizations are coupled using the stability-preserving,
second-order, time-accurate, implicit-explicit fluid-structure staggered solution procedure presented
in \cite{farhat2010robust} and the fluid-structure coupling time-step is set to $\Delta t_{F/S} = 10^{-5}$~s.

\begin{figure}[ht]
  \centering
  \includegraphics[width=0.38\textwidth]{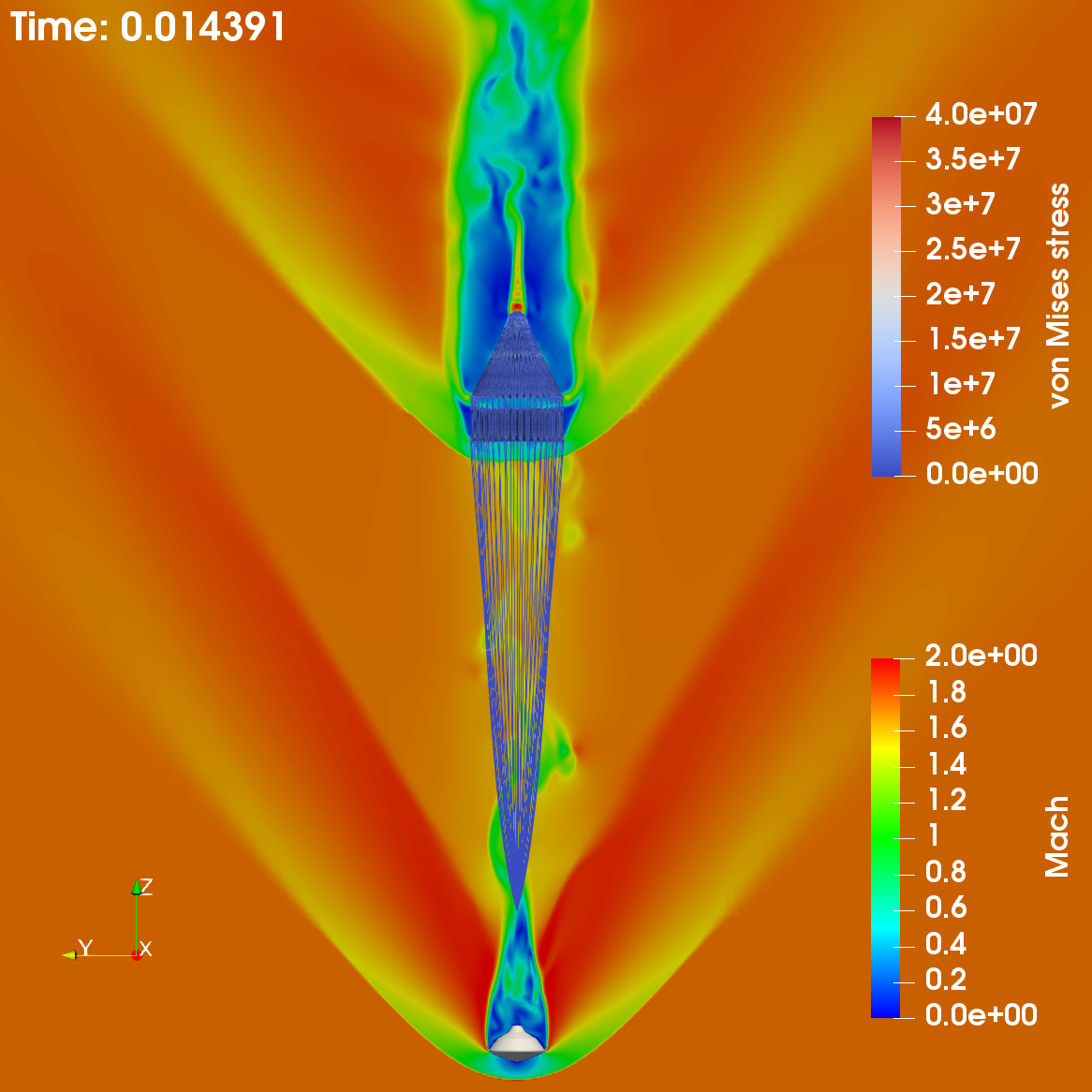}
  \includegraphics[width=0.38\textwidth]{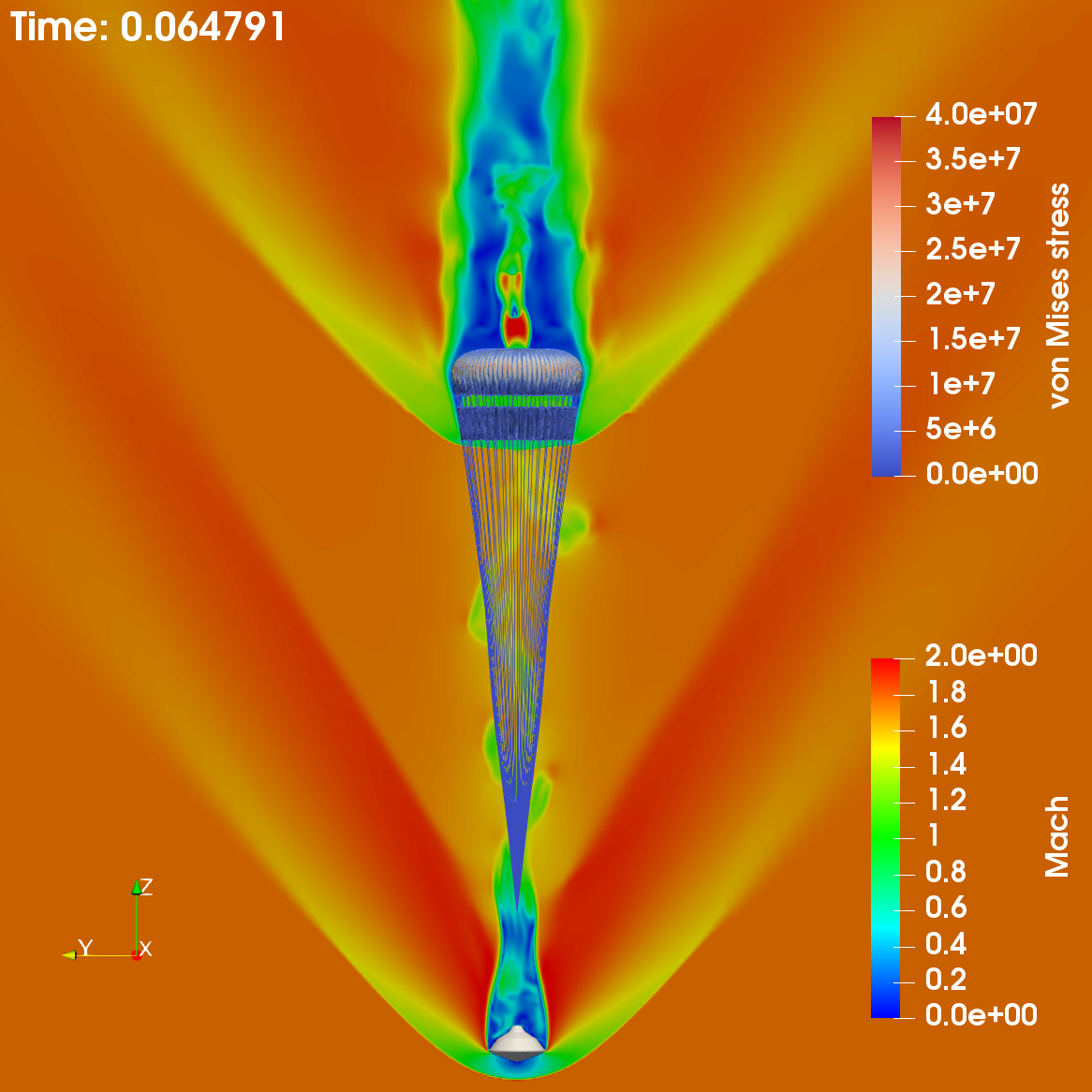}\\
\vspace{5pt}
  \includegraphics[width=0.38\textwidth]{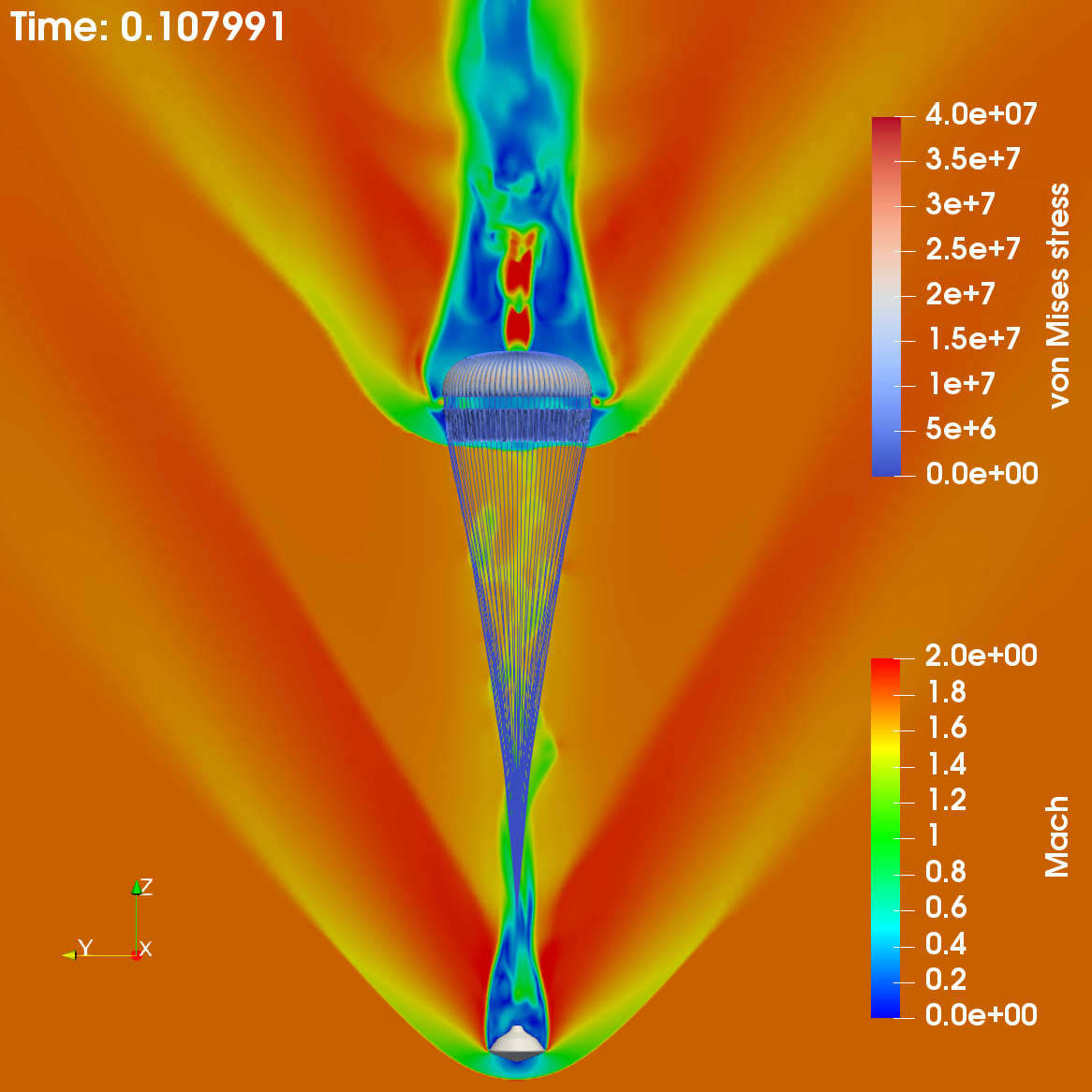}
  \includegraphics[width=0.38\textwidth]{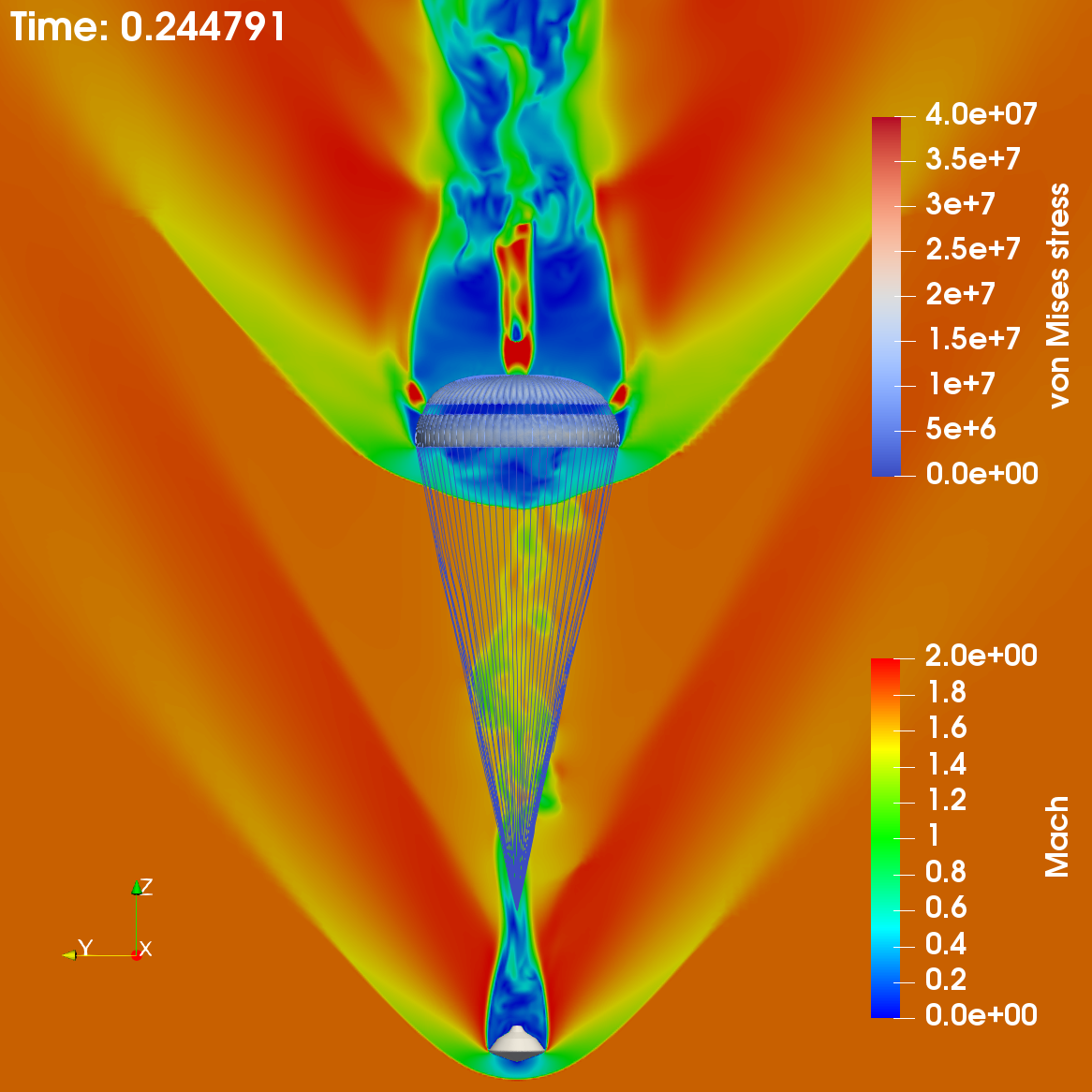}\\
\vspace{5pt}
  \includegraphics[width=0.38\textwidth]{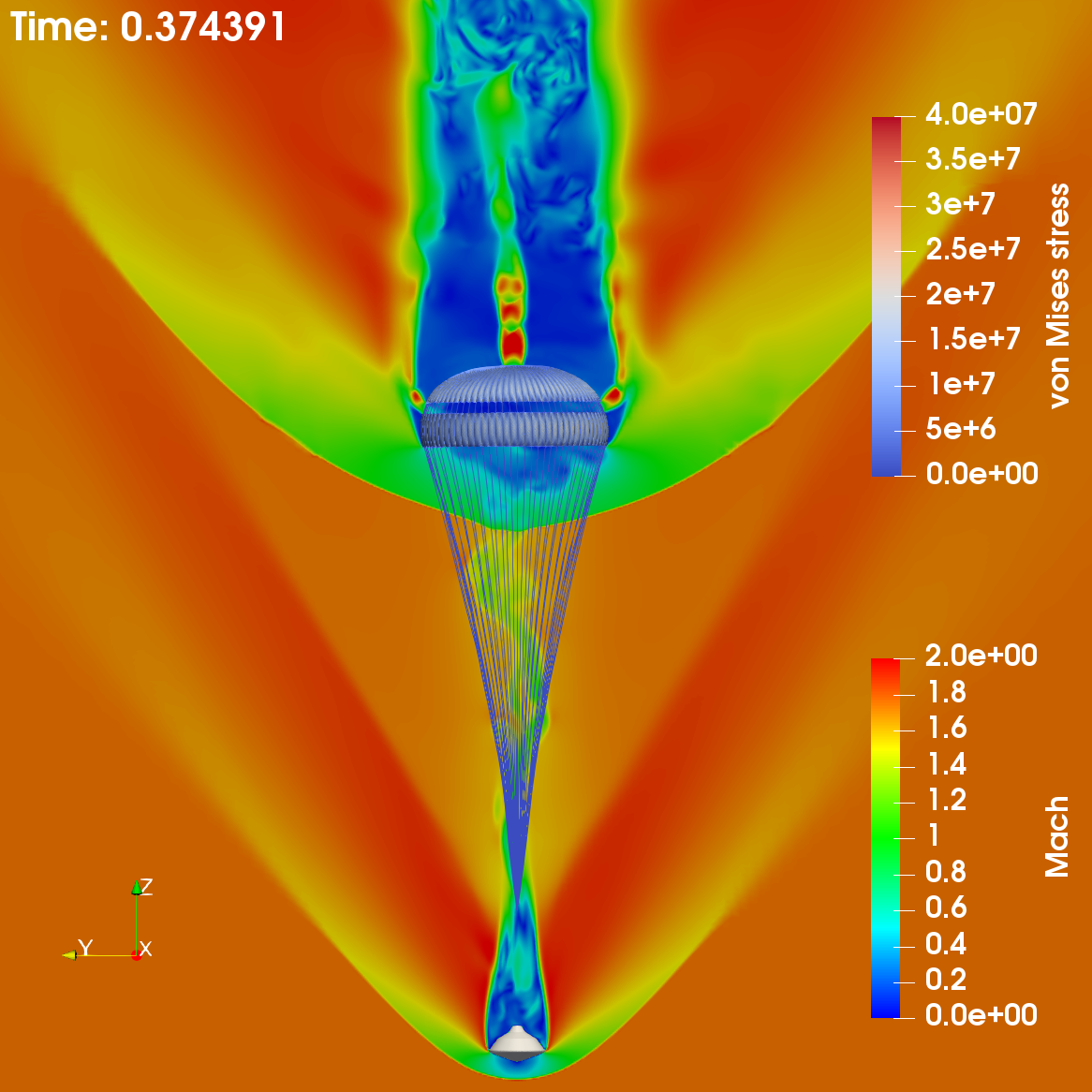}
  \includegraphics[width=0.38\textwidth]{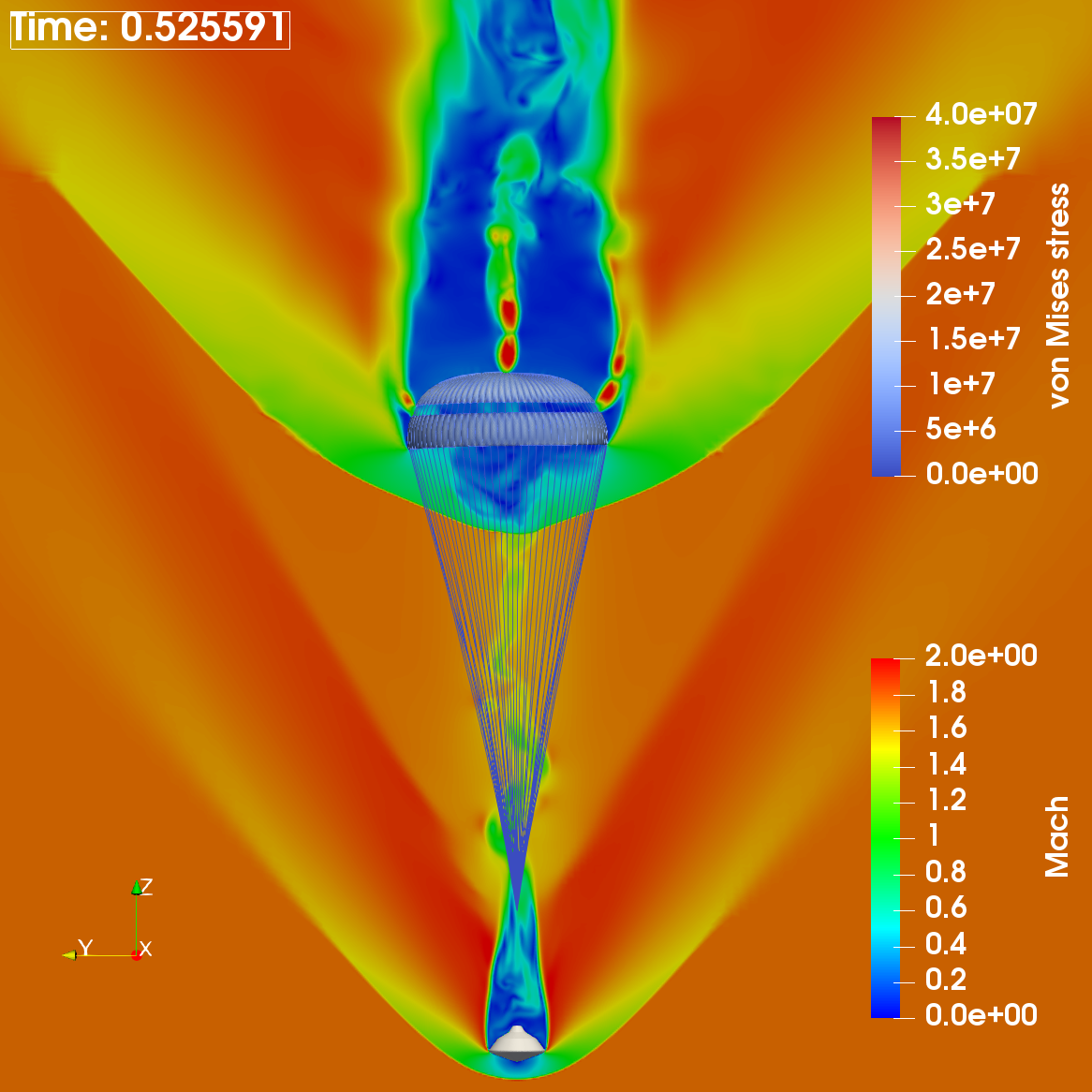}
  \caption{Simulation of the supersonic parachute inflation dynamics of a DGB system: time-evolutions of the structural system and Mach number field during deployment.}
  \label{fig:pid_mach}
\end{figure}

Figure \ref{fig:pid_mach} graphically depicts the time-evolutions of the dynamic inflation of the DGB
parachute and the flow Mach number around it. The parachute is fully inflated at approximately $t = 0.24$~s;
after this time, it starts the breathing cycles expected from a violent, high-speed, dynamic, inflation
process. 

\clearpage

\begin{figure}[ht]
  \centering
  \includegraphics[width=1.0\textwidth]{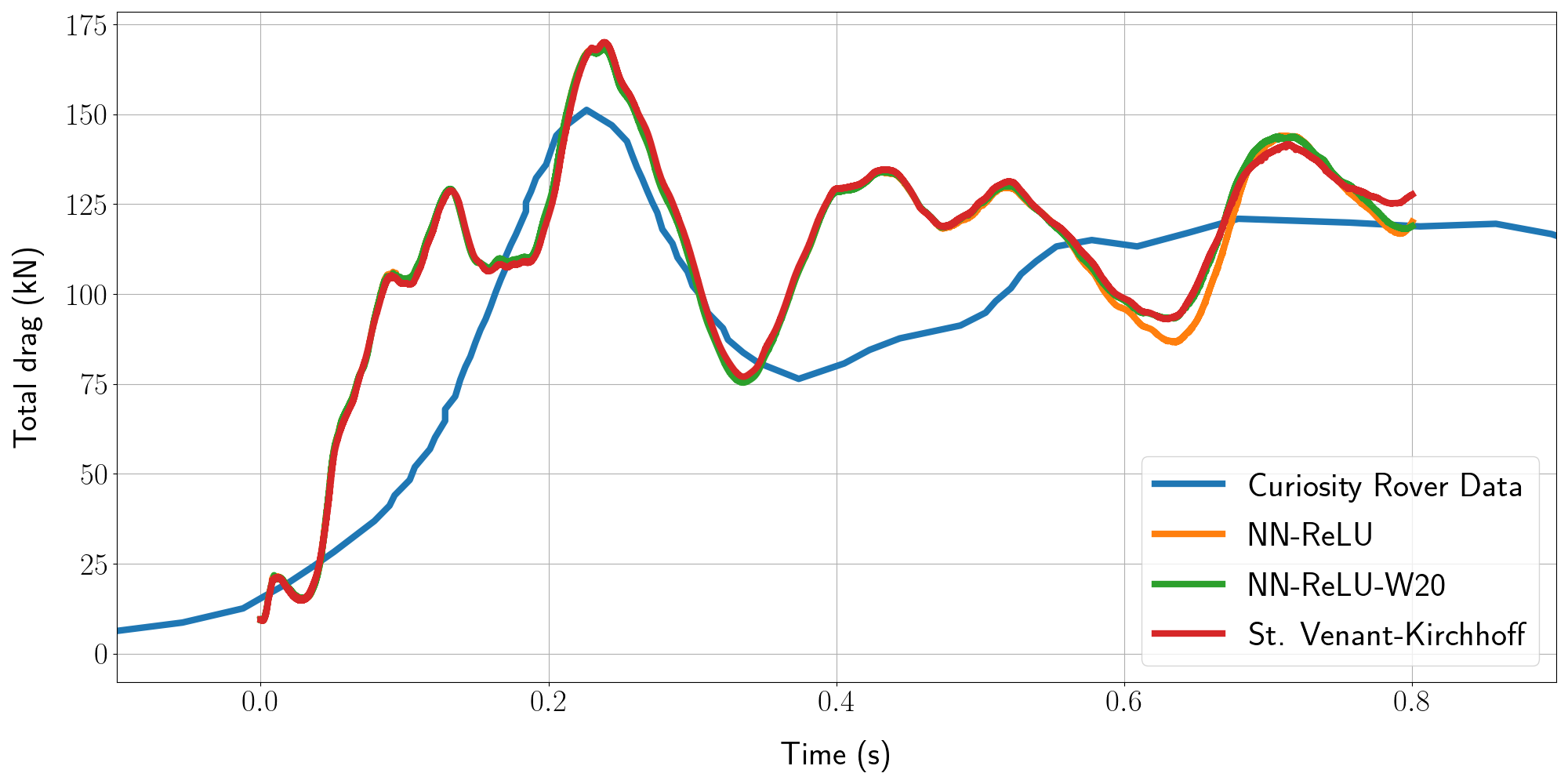}
  \caption{Simulation of supersonic the parachute inflation dynamics of a NASA DGB system: time-history of the total drag recorded during the Mars landing of Curiosity \cite{cruz2014reconstruction} 
	(blue); counterpart time-histories obtained using the coupled, multiscale, fluid-structure simulation equipped with the NN-ReLU (orange) and NN-ReLU-W20 (green) surrogate microscale models; and 
	counterpart time-history obtained using the same simulation but equipped with the St. Venant-Kirchhoff (red) constitutive model.}
  \label{fig:pid_drag}
\end{figure}

Figure \ref{fig:pid_drag} reports the time-histories of the total drag force predicted by the coupled, fluid-structure simulations described above. For validation and reference, respectively, 
this figure includes the measured time-history of the total drag generated by the parachute system during the Mars landing of NASA's rover Curiosity \cite{cruz2014reconstruction} and its counterpart
predicted using the same aforementioned simulation but equipped with the classical St. Venant-Kirchhoff model \cite{huang2020modeling}. The reader can observe that the NN-ReLU and NN-ReLU-W20
surrogate microscale models deliver stable results that are in reasonably good agreement with the flight data. The effect on drag performance of the constitutive relation used to model the woven
nylon fabric material is found to be weak. 

\begin{figure}[ht]
  \centering
  \includegraphics[width=1.0\textwidth]{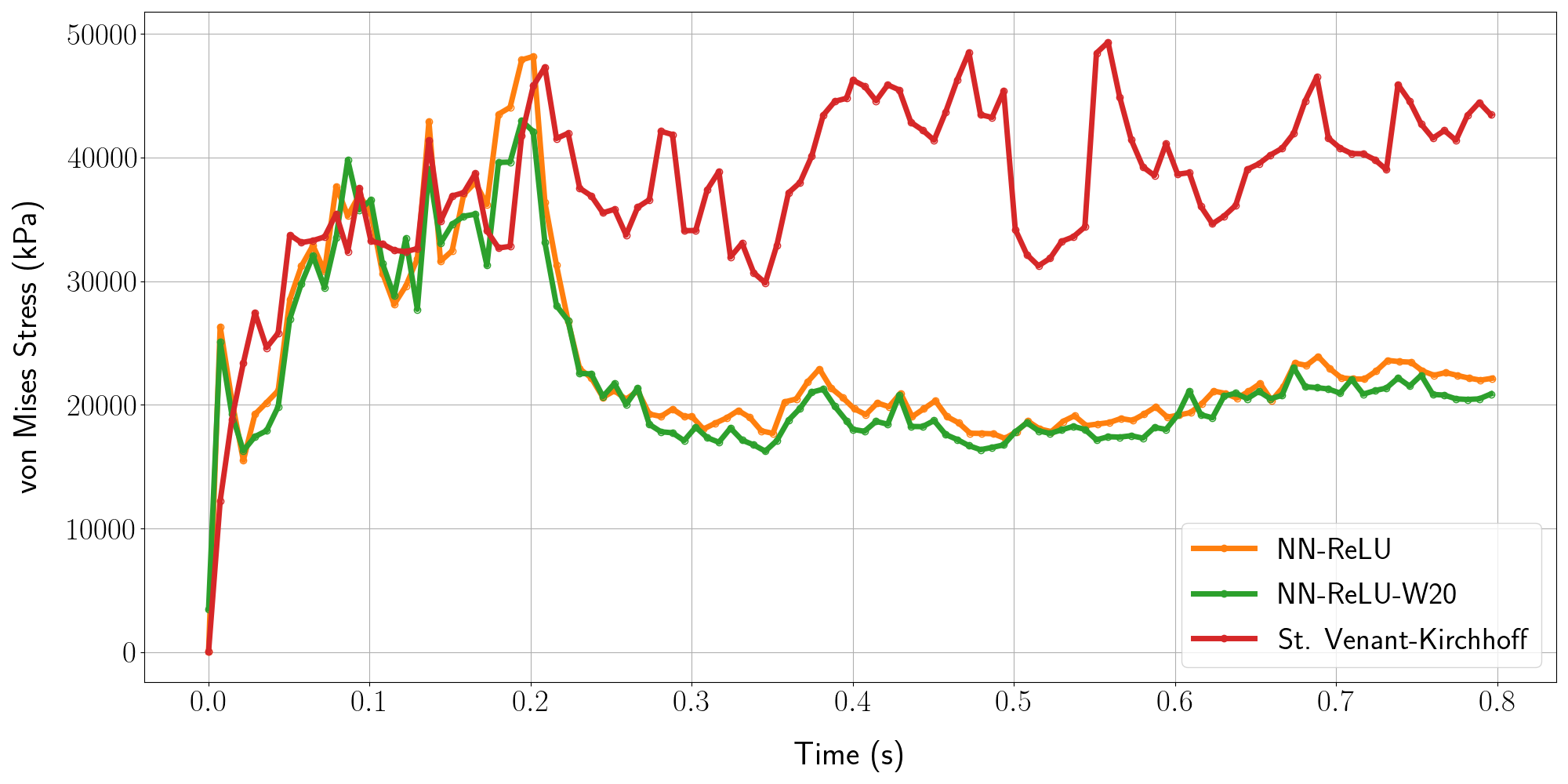}
  \caption{Simulation of the supersonic parachute inflation dynamics of a NASA DGB system: time-histories of the maximum von Mises stresses predicted by the coupled, fluid-structure interaction
	simulations equipped with the NN-ReLU (orange) surrogate microscale model, the NN-ReLU-W20 (green) counterpart model and the St. Venant-Kirchhoff (red) constitutive model.}
  \label{fig:pid_vms}
\end{figure}

\begin{figure}[ht]
  \centering
  \includegraphics[width=0.33\textwidth]{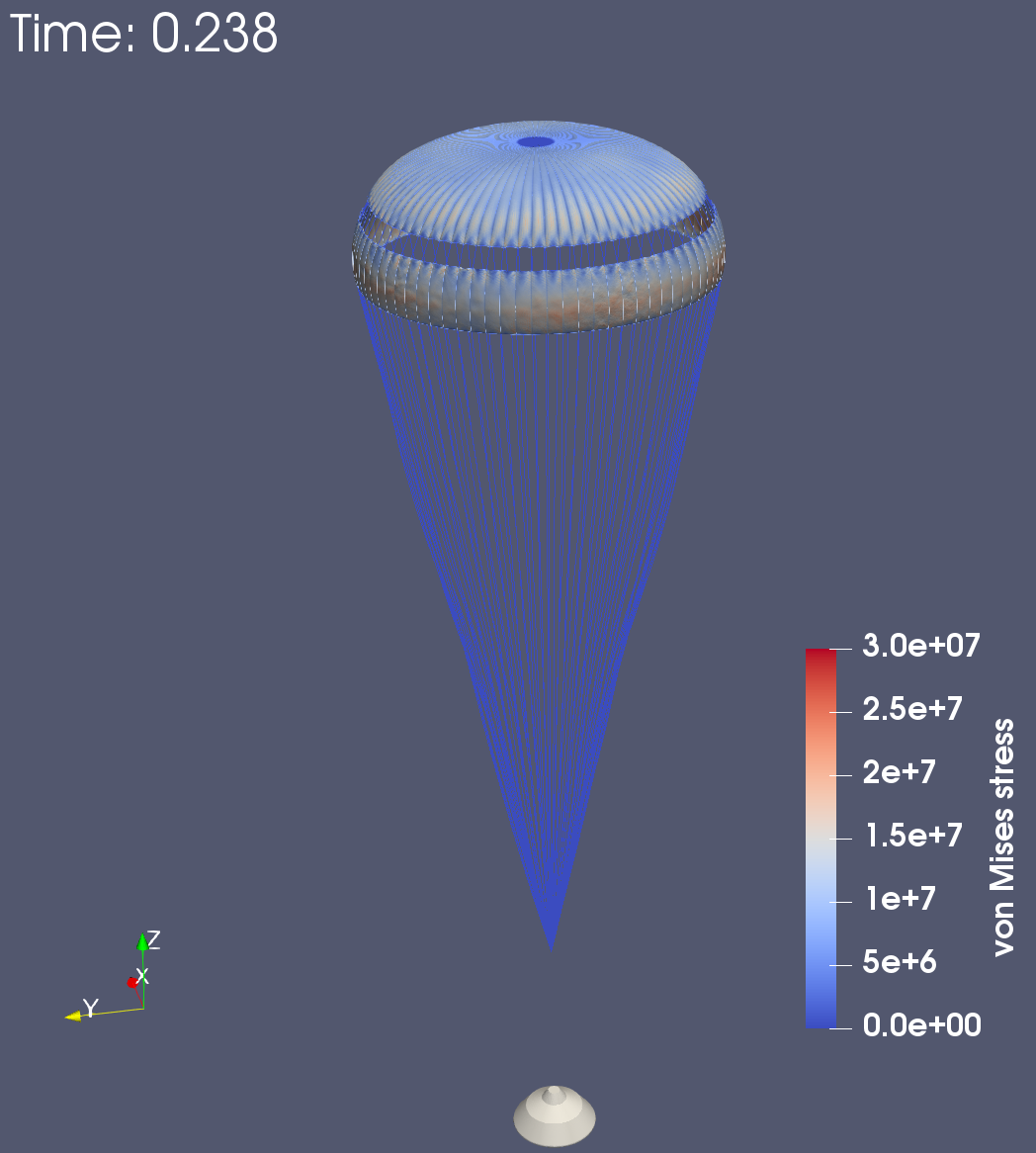}
  \includegraphics[width=0.33\textwidth]{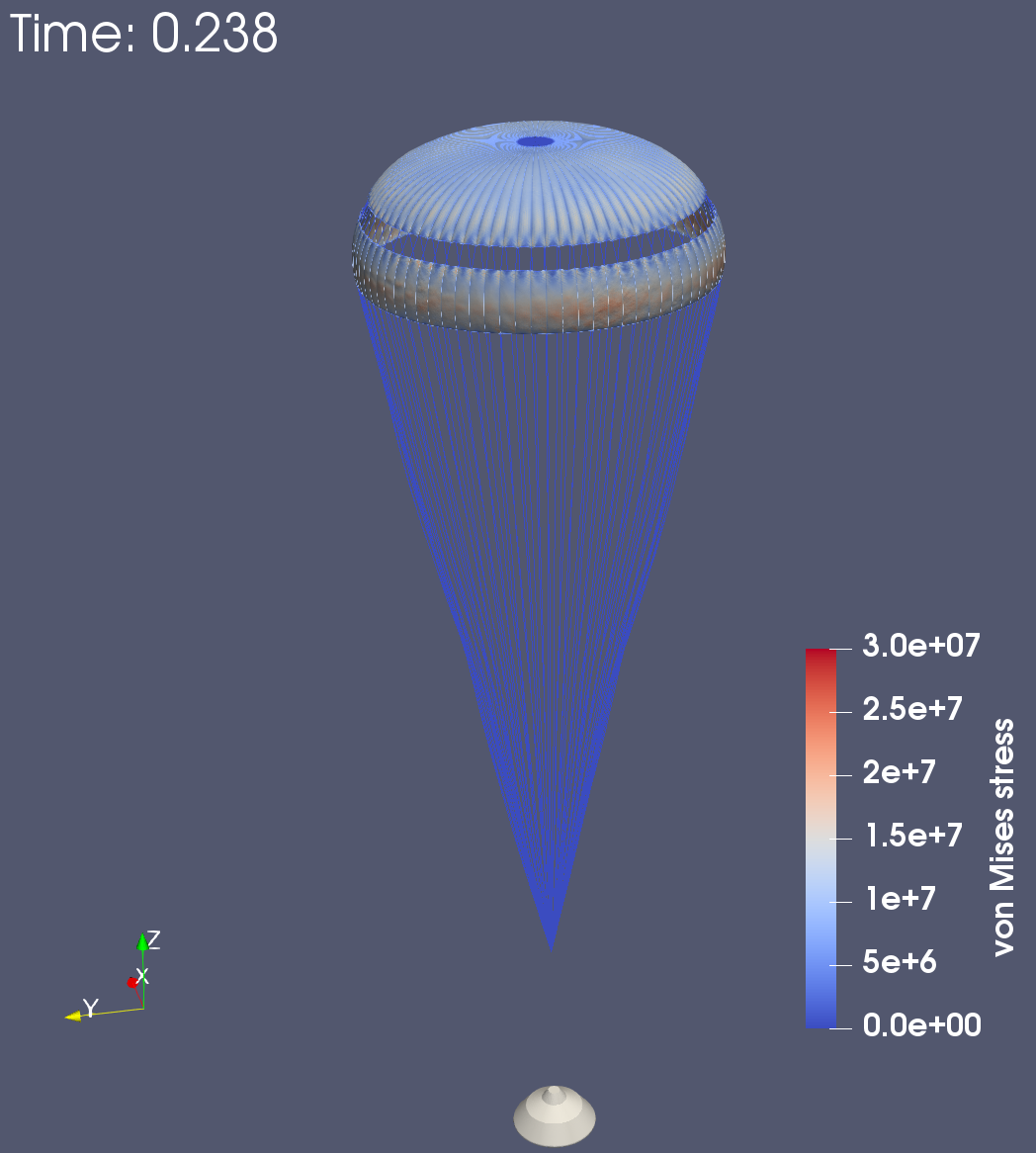}
  \includegraphics[width=0.33\textwidth]{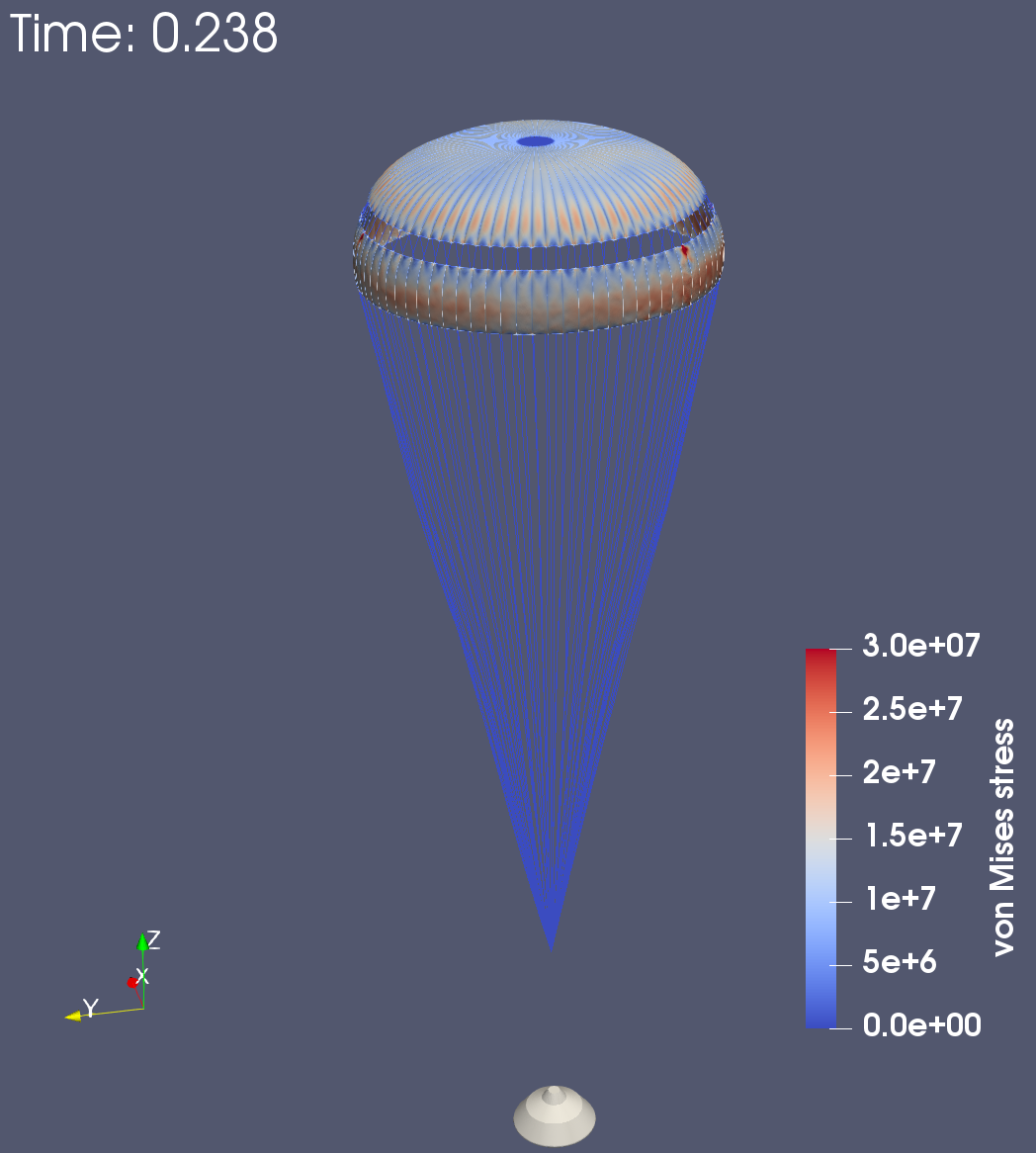}\\
\vspace{5pt}
  \includegraphics[width=0.33\textwidth]{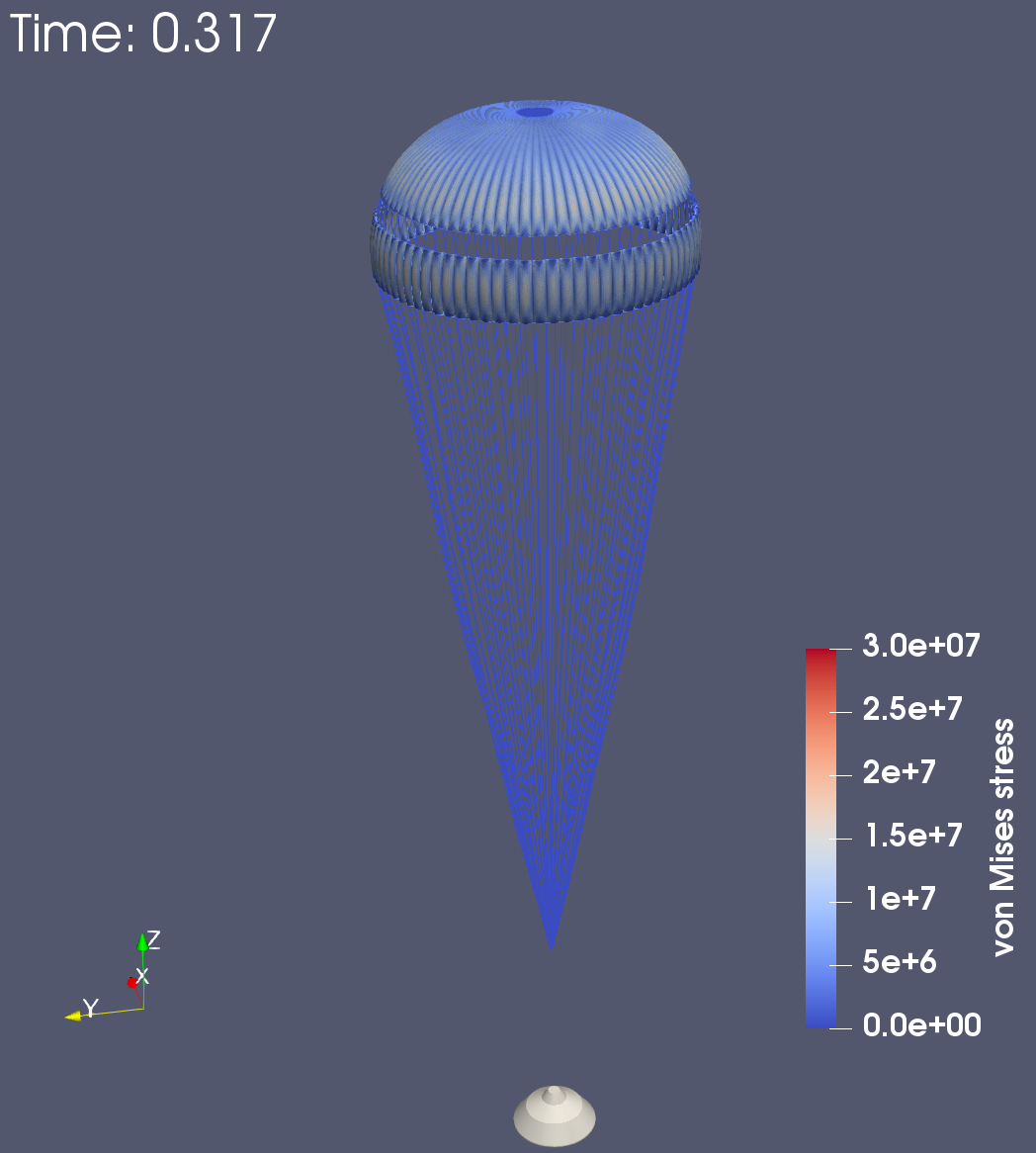}
  \includegraphics[width=0.33\textwidth]{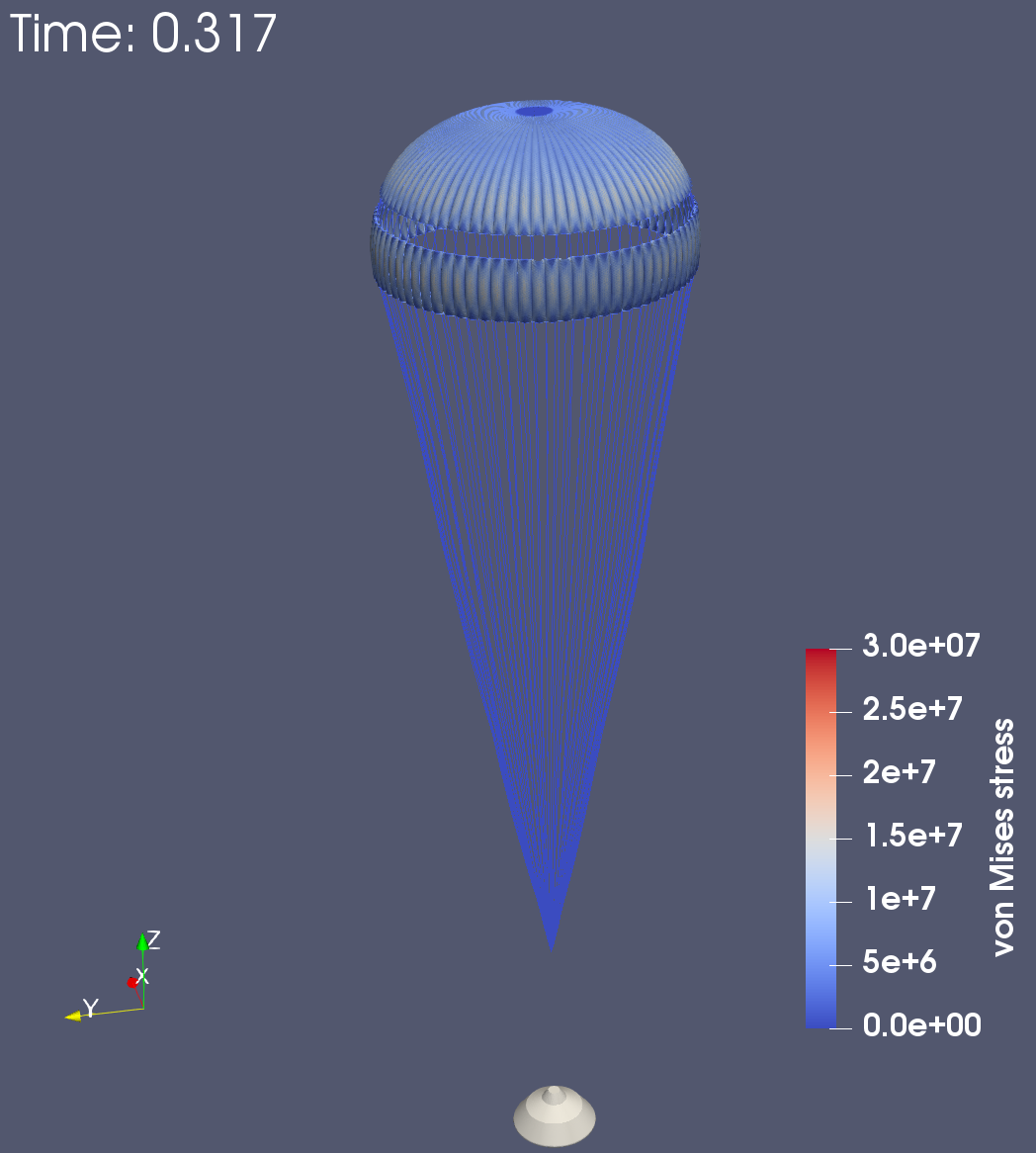}
  \includegraphics[width=0.33\textwidth]{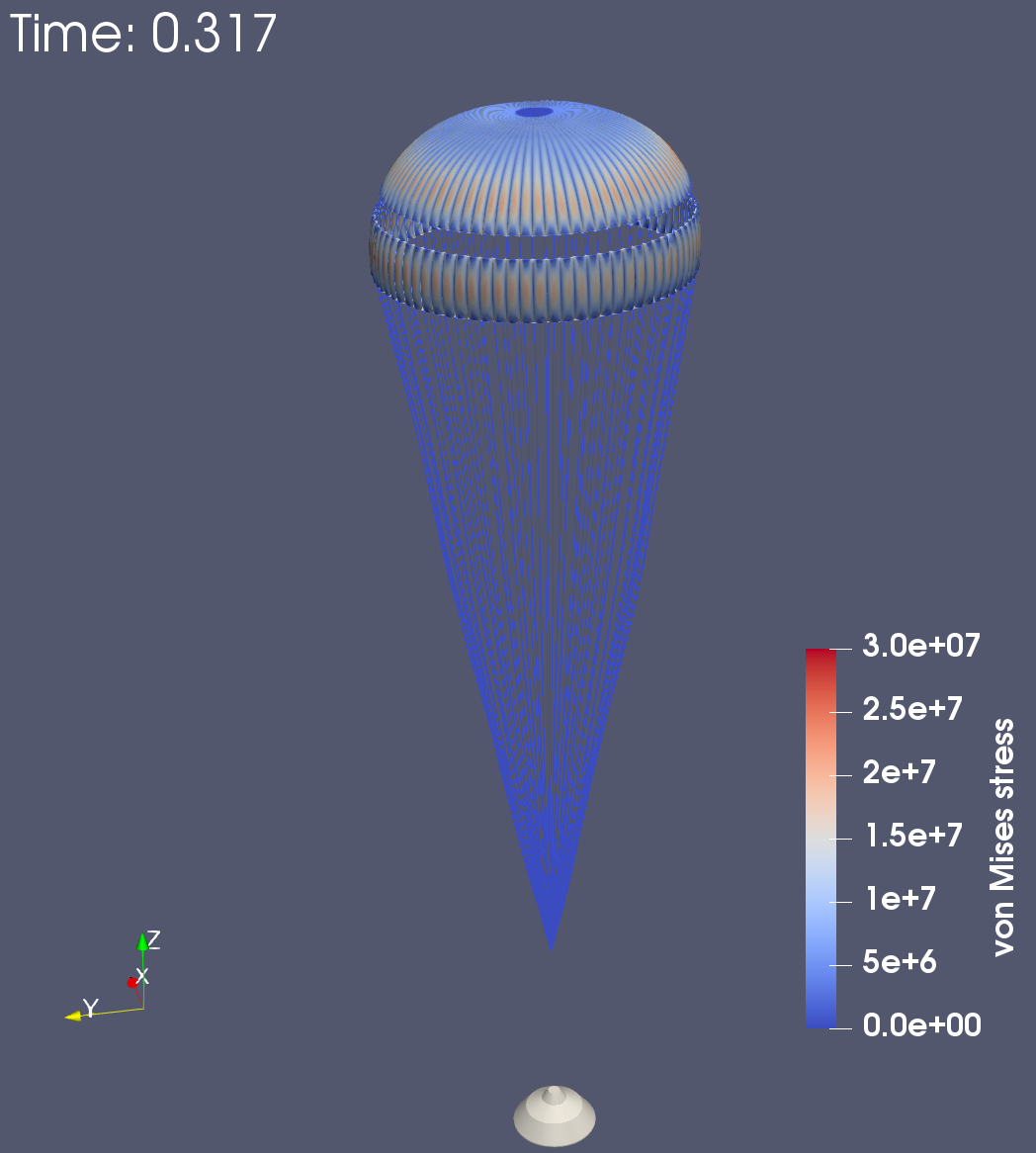}\\
\vspace{5pt}
  \includegraphics[width=0.33\textwidth]{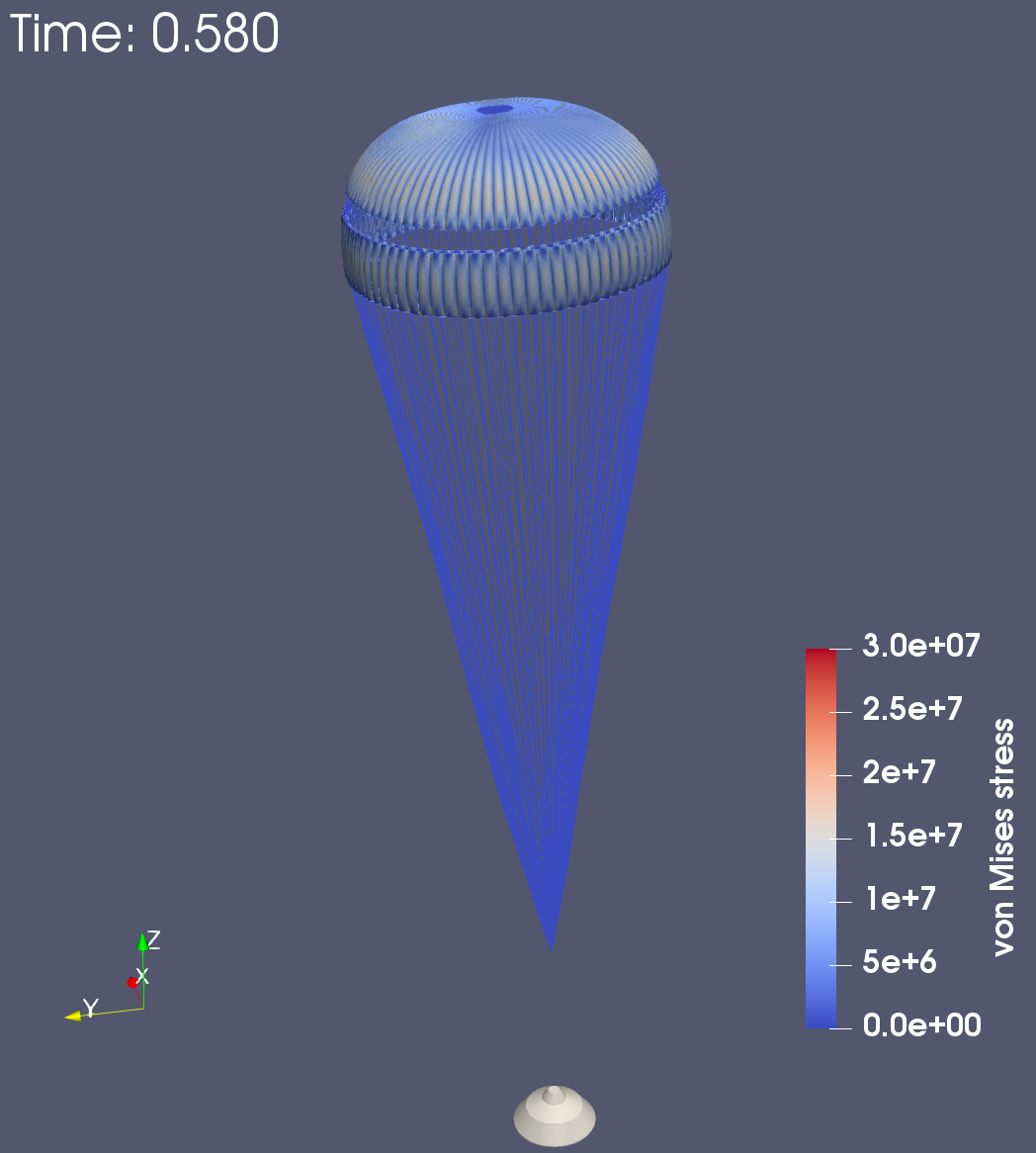}
  \includegraphics[width=0.33\textwidth]{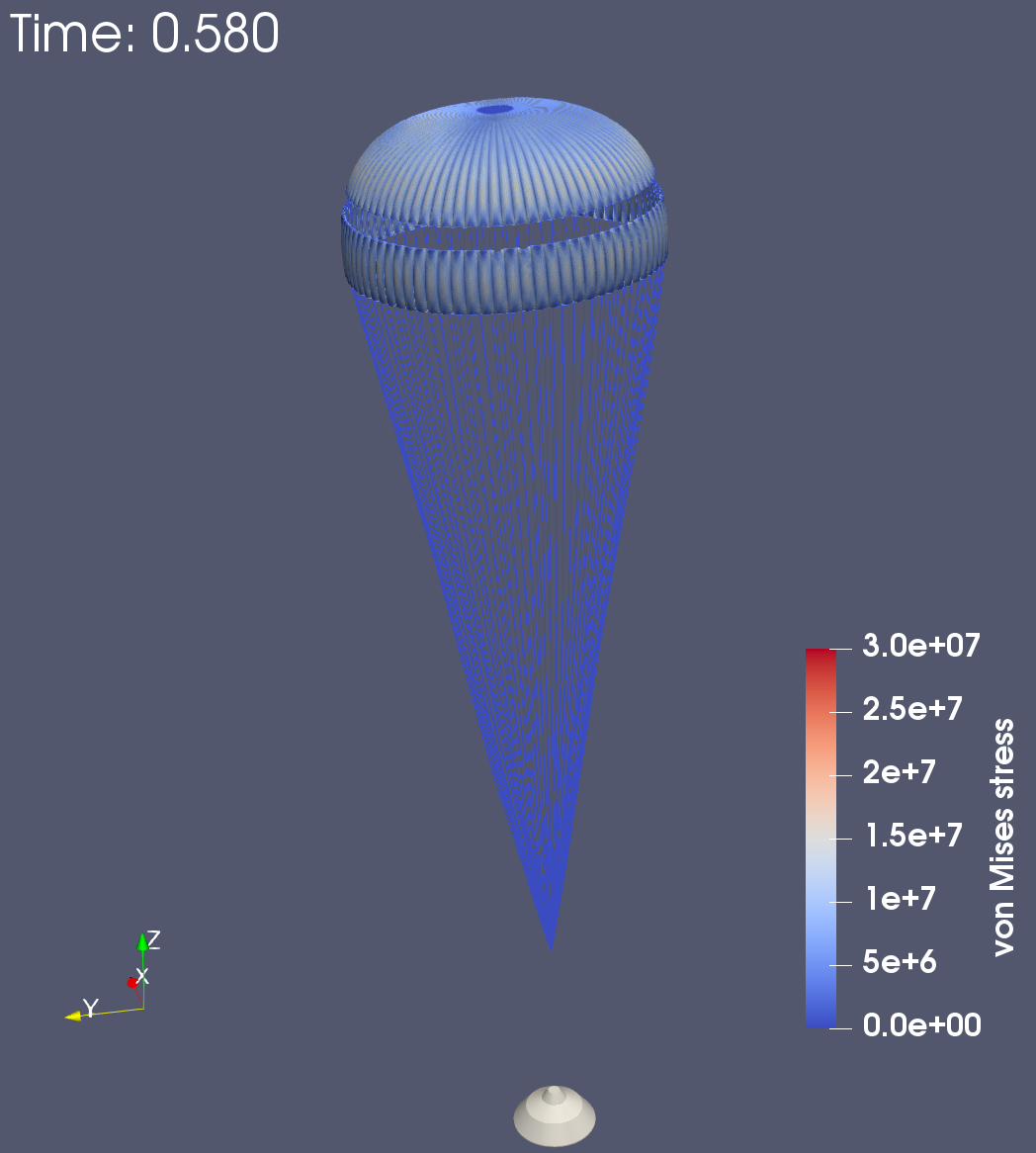}
  \includegraphics[width=0.33\textwidth]{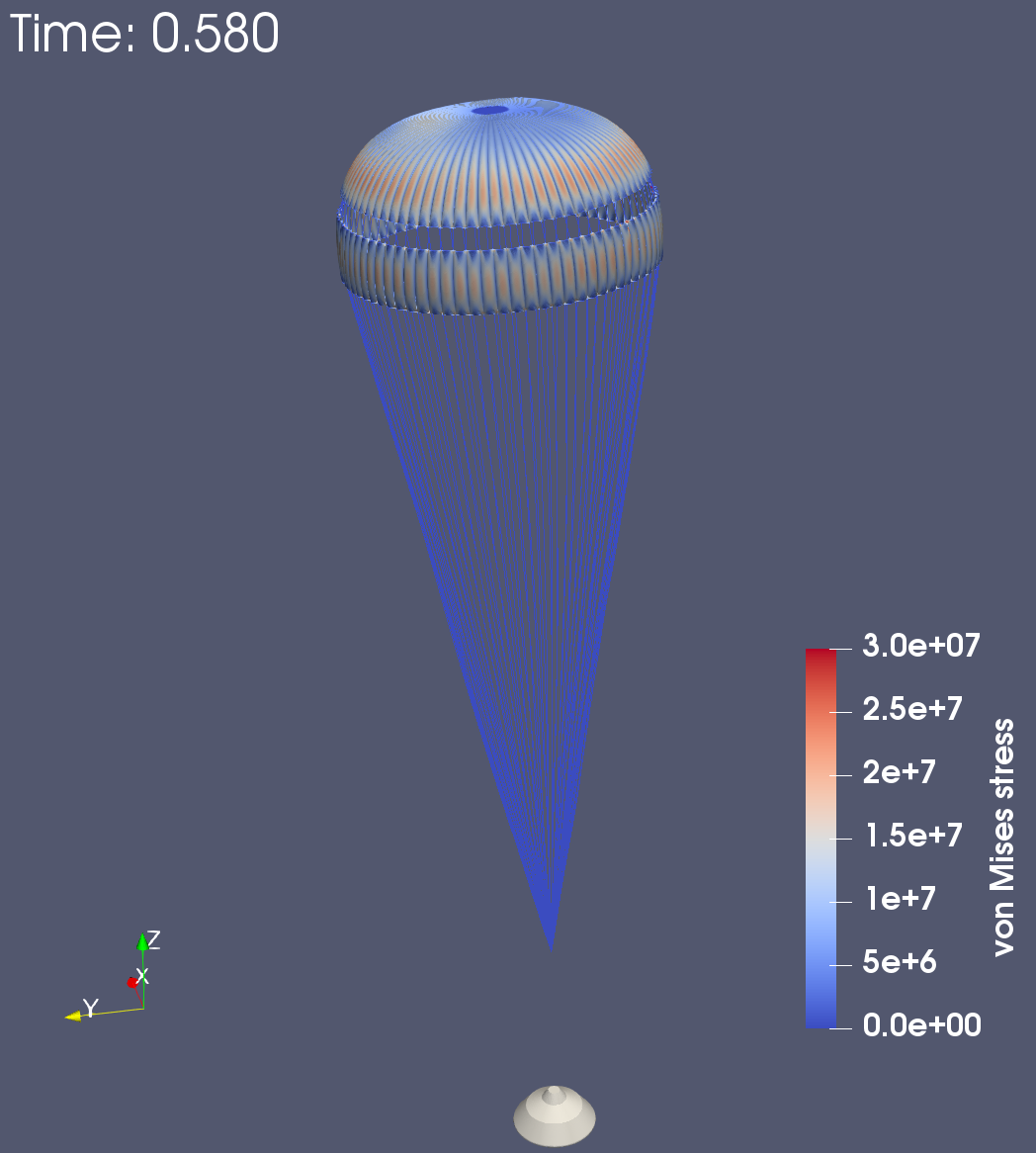}
  \caption{Simulation of the supersonic parachute inflation dynamics of a NASA DGB system: time-histories of the von Mises stress field predicted by the coupled, fluid-structure interaction 
	simulations equipped with the NN-ReLU (left) surrogate microscale model, the NN-ReLU-W20 counterpart model (middle) and the St. Venant-Kirchhoff (right) constitutive model.}
  \label{fig:pid_vms_te}
\end{figure}

Figure \ref{fig:pid_vms} reports the time-histories of the maximum von Mises stresses -- an indicator of material failure -- predicted by the aforementioned coupled, fluid-structure interaction
simulations. Similar stress results are delivered by the NN-ReLU and NN-ReLU-W20 surrogate microscale models, which indicates that for this application, the shear effect of the woven nylon fabric is not 
significant.  However, the results delivered by the St. Venant-Kirchhoff model show that the flexibility with respect to shearing and compression of the multiscale woven fabric model highlighted in 
Section \ref{sec:sst} leads to lower von Mises stresses in the parachute breathing cycle, after full inflation. This disparity between the results obtained using the classical St. Venant-Kirchhoff and 
multiscale models is also highlighted in Figure \ref{fig:pid_vms_te}, which depicts the time-evolutions of the entire von Mises stress fields. Although further (experimental) investigation is required 
to conclude which model is more reliable, this comparison illustrates the potential of a multiscale constitutive model for improving the prediction of material failure. 

The combined offline/online computational costs of all coupled, fluid-structure interaction simulations discussed above are reported in Table \ref{tab:time}, component-by-component. It is worth
mentioning that the {\it estimated} simulation time for the direct FE$^2$ simulation -- based on the multiplication of the number of FE$^2$ model evaluations on each CPU and the cost of a single 
evaluation -- is also reported in Table \ref{tab:time}: it suggests that the NN-based surrogate microscale models lead to speed up factors of the order
of $7\times10^4$ (all training costs included), which demonstrates the potential of NN-based surrogate microscale models for constitutive modeling.

\begin{table}
\centering
\begin{tabular}{c|c|c}
\toprule
& Number of cores & Wall-clock time (hrs) \\
\midrule
Training data generation & 1 & 40 \\
\midrule
Training & 1 & 0.01 \\
\midrule
Flow solver & 480 & 96.14 \\
\midrule
\shortstack{Structure solver \\
	with FE$^2$ + NN-ReLU} & 96 & 19.52 \\
\midrule
\shortstack{Structure solver \\
	  with FE$^2$ + NN-ReLU-W20} & 96 & 24.82 \\
\midrule
	\shortstack{Structure solver \\with St. Venant-Kirchhoff} & 96 & 2.96 \\
\midrule
	\shortstack{Structure solver\\FE$^2$ (estimated)} & 96 & 4,204,336.55
\\
\bottomrule
\end{tabular}
\caption{Simulation of the supersonic parachute inflation dynamics of a NASA DGB system: computational costs.}
\label{tab:time}
\end{table}

\clearpage

\section{Conclusions}
\label{sec:conclude}

The general framework for computationally tractable, nonlinear, multiscale modeling of membrane, woven fabrics presented in this paper is
enabled by the coherent exploitation of several key, established pillar methodologies:
\begin{itemize}[topsep=0pt] 
\itemsep-0.35em
\item The computational homogenization approach known as finite element squared (FE$^2$) based on the concept of
      a locally attached microstructure.
\item The numerical enforcement of the membrane's plane stress condition.
\item The ``discovery'' of a surrogate, microscale model such as an artificial neural network (NN)-based regression model
	using data generated by many multiscale numerical simulations of the behavior of a small woven fabric coupon.
\item The acceleration of the training of the above NN using nonlinear projection-based model order reduction (PMOR) and hyperreduction.
\end{itemize}

The proposed computational framework encompasses a cascade of multiscale models, ranging from the highest fidelity (without any
surrogate model) to the lowest (linear regression surrogate model). The proposed discovery at the finest scale of a surrogate constitutive model by means
of numerical coupon testing is analogous to the experimental testing procedure used to identify the parameters (e.g. Young's modulus or
Poisson's ratio) of conventional material models. A highlight of the overall approach is that while experimental data is typically limited to 
uniaxial tension (occasionally biaxial and/or shear data may also be available), numerical data suffers from no such limitation. Using the concept
of numerical coupon testing, an entire parameter space of physically admissible combinations of normal and shear strains can be explored in order 
to characterize complex and unconventional materials, and support the discovery of a constitutive law. Furthermore, using PMOR equipped
with hyperreduction (see Appendix \ref{sec:micro-rom}), the aforementioned exploration of a large parameter space can be performed in a multiscale
setting in practical wall-clock time. PMOR continues to be an active and fertile area of research that can be leveraged to extend and improve this
framework. In particular, the recent emergence of in-situ training methodologies \cite{he2020situ} presents an attractive option to streamline and
enhance PMOR utilization by eliminating the conventional and potentially cumbersome offline-online decomposition of computational effort and
vulnerabilities associated with extrapolation.

All of the above conclusions are supported in this paper by the successful demonstration of the proposed computational framework for
the simulation of a supersonic parachute inflation dynamics problem in Martian atmospheric entry conditions, for which flight data is available. 
For this application, a NN-based surrogate microscale model is constructed and trained in a large parameter space in 40 hours wall-clock time using a single 
computational core. This surrogate model is shown to enable the proposed overall nonlinear multiscale framework to achieve computational tractability. Specifically,
the coupled, multiscale, fluid-structure interaction simulation of the supersonic, dynamic inflation process of the parachute and a few of its breathing
cycles is completed in about 116 hours wall-clock time (less than 5 days) on 576 cores of a Linux cluster (82\% of this wall-clock time is consumed by
the computation of the turbulent flow). Particularly, the NN-based surrogate microscale model is shown to reduce by almost five orders of magnitude the wall-clock time
that would otherwise be required for performing the multiscale structural dynamics computations within the same fluid-structure interaction simulation using only the proposed tailoring of the FE$^2$ 
framework to membrane woven fabrics. Equally importantly, the time-history of the total drag force predicted using the proposed computational framework is found to match well its 
flight-recorded counterpart.

\section*{Acknowledgments}
Philip Avery, Daniel Huang, Johanna Ehlers and Charbel Farhat acknowledge partial support by the Jet
Propulsion Laboratory (JPL) under Contract JPL-RSA No. 1590208 and partial support by the National Aeronautics
and Space Administration (NASA) under Early Stage Innovations (ESI) Grant NASA-NNX17AD02G. Parts of this
work were completed at the JPL, California Institute of Technology, under a contract with NASA. Optical
microscope photographs were taken by Cheyenne Hua at the JPL. Any opinions, findings and conclusions
or recommendations expressed in this paper are those of the authors and do not necessarily reflect the
views of JPL or NASA.

\section*{Data Availability}
The data that support the findings of this study are openly available in the repository at \url{https://github.com/Zhengyu-Huang/Fabric-Data.git}.

\appendix
\section{Multiscale projection-based model order reduction approach for space exploration}
\label{sec:micro-rom}

For the sake of completeness, a nonlinear, multiscale, projection-based model order reduction (PMOR)/hyperreduction
framework is presented here for dramatically accelerating the training of a regression-based artificial neural network (NN) 
in a large parameter space, in view of using it as a surrogate microscale model. The described approach constitutes a generalization of 
the framework first presented in \cite{zahr2017multilevel} to:
\begin{itemize}
\item Include a treatment of contact based on the PMOR method originally proposed in \cite{balajewicz2016projection}, which features a non-negative matrix factorization scheme for
	the construction of a positive reduced-order basis (ROB) for the contact forces.  
\item Accommodate a novel training strategy based on the concept of a coupon test analogy introduced in Section \ref{sec:micro-surrogate}.
\end{itemize}

For the the sake of simplicity and clarity, but without any loss of generality, the proposed PMOR approach is described here only
for the microscale level of a two-scale (macro-micro) model. Specifically, Proper Orthogonal Decomposition (POD) is used to construct a 
projection-based reduced-order model (PROM) at the microscale level and a computational approach based on the energy conserving sampling and weighting (ECSW) 
method \cite{farhat2014dimensional, farhat2015structure} is used to hyperreduce the constructed PROM. Training is performed offline (i.e. {\it a priori}) using a small, {\it multiscale
coupon model}.

\subsection{Reduction of the primal unknowns}
\label{subsec:prom}
At the microscale (scale $1$), the number of primal dofs $n_1$ of the computational model is reduced by searching for the primal solution $\ubm_1$ of the typical microscale
problem in a carefully constructed low-dimensional subspace, i.e.,
\begin{equation}
\label{eqn:ansatz}
  \ubm_1 \approx \Vbm_1\ybm_1
\end{equation}
where $\Vbm_1 \in \Rbb^{n_1 \times r_1}$ is a ROB representing a low-dimensional
subspace, $\ybm_1 \in \Rbb^{r_1}$ is the vector of generalized coordinates of $\ubm_1$ in this basis
and $r_1 \ll n_1$. The ROB is chosen to be orthonormal with respect to the identity matrix, i.e.,
\begin{equation*}
  \Vbm_1^T\Vbm_1 = \Ibm
\end{equation*}

As mentioned above, the ROB is constructed using POD and the method of snapshots \cite{sirovich1987turbulence}.
To this end, $m_1$ solution snapshots of (\ref{eqn:fem1}), $\{\ubm_1^{(1)}, \dots, \ubm_1^{(m_1)}\}$, are computed at scale $1$
for different prescribed boundary displacements and collected in the {\it primal} snapshot matrix 
\begin{equation*}
  \Ybm_1^u = \begin{bmatrix} \ubm_1^{(1)} & \cdots & \ubm_1^{(m_1)} \end{bmatrix}
\end{equation*}
Then, this matrix is compressed using the singular value decomposition (SVD) method and $\Vbm_1$ is constructed using the first $r_1$ singular vectors of $\Ybm_1^u$,
where $r_1$ is determined from the application of a retention criterion to the energy of the singular values.

From (\ref{eqn:micro-deform-ansatz}), it follows that the constrained dofs of the microscale displacement vector $\mathring{\ubm}_1$ lies in a low-dimensional subspace associated with a vector of
generalized coordinates identified as the column-wise vectorization of the right stretch strain tensor
$\Ubm_0 - \Ibm$, i.e.
\begin{equation*}
\label{eqn:ansatzc}
\begin{aligned}
  \mathring{\ubm}_1 &= \Pibold_1 \begin{bmatrix} \mathring{\Xbm}_1 & 0 & 0 \\
                                                 0 & \mathring{\Xbm}_1 & 0 \\
                                                 0 & 0 & \mathring{\Xbm} \end{bmatrix} \mathrm{vec} \left(\Ubm_0 - \Ibm\right) &
                    &= \mathring{\Vbm}_1 \mathring{\ybm}_1
\end{aligned}
\end{equation*}
where $\Pibold_1$ is a permutation matrix and $\mathring{\Xbm}_1$ is a matrix whose three columns represent
the $x$, $y$ and $z$ nodal coordinates, respectively, of the constrained nodes located on the boundary of the microscale model.
The definition of $\bar{\ybm}_1$ follows from to the notational convention (\ref{eqn:notation}). Hence, a basis encompassing both unconstrained and constrained dofs can be represented, 
up to a permutation, as
\begin{equation*}
  \overline{\Vbm}_1 = \begin{bmatrix} \Vbm_1 & 0 \\
                                      0 & \mathring{\Vbm}_1 \end{bmatrix}
\end{equation*}

The dimensionality of the discrete governing equations (\ref{eqn:fem1}) is reduced at scale $1$ by
performing a Galerkin projection, i.e., substituting (\ref{eqn:ansatz}) in these equations and projecting
the first of them onto the column space of $\Vbm_1$. This leads to the PROM
\begin{subequations}
\label{eqn:rom1}
\begin{align}
  \Vbm_1^T\fbm_1^{int}\left(\overline\Vbm_1\bar\ybm_1\right) + \Vbm_1^T\Gbm_1\left(\overline\Vbm_1\bar\ybm_1\right)\lambdabold_1 &= 0 \\
  \gbm_1\left(\overline\Vbm_1\bar\ybm_1\right) &\ge 0 \\
  \lambdabold_1 &\le 0 \\
  \lambdabold_1^T \gbm_1\left(\overline\Vbm_1\bar\ybm_1\right) &= 0
\end{align}
\end{subequations}

Despite the fact that the equations (\ref{eqn:rom1}) are characterized by a reduced dimensionality, their
solution remains computationally intensive due to the presence of the nonlinear term $\fbm_1^{int}$.
Indeed, the projection of this term implies that every evaluation of $\Vbm_1^T\fbm_1^{int}$ requires
the reconstruction of the full state using the approximation $\overline\Vbm_1\bar\ybm_1$, the integration
and assembly of the internal force vector over the entire computational mesh and its projection onto
the subspace represented by the ROB $\Vbm_1$. Because such computations scale with the size $n_1$ of
the high-dimensional model at level $1$, they cannot be performed using limited resources or at low computational
cost -- and much less in real time. Hence, they constitute a substantial bottleneck in the solution of
(\ref{eqn:rom1}). For this reason, a number of hyperreduction methods have been proposed to overcome
this bottleneck introduced by nonlinear terms. For solid mechanics and structural dynamics problems,
the ECSW method is preferred due to its desirable structure-preserving and
numerical stability properties \cite{farhat2015structure}. However, any other efficient hyperreduction
method can be equally used, in principle, to overcome the aforementioned computational bottleneck.

As introduced in \cite{farhat2014dimensional}, the ECSW method amounts to a ``mesh reduction and quadrature'' algorithm
which samples a set of mesh elements $\Vcal_1^{\prime} \subset \Vcal_1$ and attributes to each sampled element
$e$ a positive weight $\alpha_1^e > 0$ such that
\begin{equation*}
\begin{aligned}
  \overline\Vbm_1^T\bar\fbm_1^{int}\left(\overline\Vbm_1\bar\ybm_1\right)
  & = \sum_{e \in \Vcal_1} \left(\overline\Vbm_1^e\right)^T \bar\fbm_1^{{int}^e}\left(\overline\Vbm_1^e\bar\ybm_1\right)
  & \\
  & \approx \sum_{e \in \Vcal_1^{\prime}} \alpha_1^e\left(\overline\Vbm_1^e\right)^T \bar\fbm_1^{{int}^e}\left(\overline\Vbm_1^e\bar\ybm_1\right)
  & = \bar\fbm_{1_r}^{int}\left(\bar\ybm_1\right)
\end{aligned}
\end{equation*}
In the above expressions, the superscript $e$ designates the \emph{restriction} of a global vector or
matrix to element $e$ and the reduced mesh $\Vcal_1^{\prime}$ ($|\Vcal_1^{\prime}| \ll |\Vcal_1|$) can be computed
using Lawson and Hanson's Non-Negative Least Squares (NNLS) algorithm \cite{lawson1995solving}, or an alternative
L1 minimization algorithm \cite{chapman2017accelerated}, in a training step that seeks to minimize the
size of $\Vcal_1^{\prime}$ while maintaining an acceptable approximation error for the ensemble of the
training data.

In addition to achieving a computational complexity that scales with the size $n_1$ of the PROM only in the computation
of the components of the internal force vector corresponding to unconstrained dofs, ECSW and its reduced mesh
ensure that scale transmission is performed efficiently, i.e., without any operation whose computational
complexity scales with $|\Vcal_1|$. This is evident in the transmission to finer scales, where ${\mathring{\ubm}}_1^e$
is required for each $e \in \Scal_1^{\prime}$; as for transmission to coarser scales, the homogenized macroscopic unsymmetric Biot stress tensor is approximated as
\begin{equation*}
  \mathrm{vec}(\Bbm_0) \approx \frac{1}{|\Bcal_1|} \mathring{\fbm}_{1_r}\left(\bar\ybm_1\right)
\end{equation*}
where $\Scal_1^{\prime} \subset \Scal_1$ denotes the subset of surface elements contained in the reduced
mesh $\Vcal_1'$ and $\mathring{\fbm}_{1_r}$ is in general the restriction of the \emph{total} vector of
reduced forces -- both internal and contact -- to the constrained generalized coordinates, i.e.
\begin{equation*}
  \mathring{\fbm}_{1_r} = \mathring{\fbm}_{1_r}^{int}\left(\bar\ybm_1\right) +
                          \mathring{\Vbm}_1^T\mathring{\Gbm}_1\left(\overline\Vbm_1\bar\ybm_1\right)\lambdabold_1
\end{equation*}
In some cases though, the mesh of the microscale model can be constructed in such a way that the contact forces will
contribute nothing to this quantity. This requires maintaining a separation of at least one element between the contact surface and the boundary of the mesh.

\subsection{Reduction of the dual unknowns}
\label{subsec:drom}
At the microscale (here, scale $1$), the number of dual dofs $n^\lambda_1$ of the computational model can also
be reduced by searching for the dual solution $\lambdabold_1$ of the problem of interest in another carefully
constructed low-dimensional subspace, i.e.,
\begin{equation}
\label{eqn:dualansatz}
  \lambdabold_1 \approx \Wbm_1\zbm_1
\end{equation}
where $\Wbm_1 \in \Rbb^{n^\lambda_1 \times r^\lambda_1}$ is a dual ROB representing a
low-dimensional subspace, $\zbm_1 \in \Rbb^{r^\lambda_1}$ is the vector of generalized coordinates of
$\lambdabold_1$ in this basis and $r^\lambda_1 \ll n^\lambda_1$. The dual ROB is chosen such that it has
no negative entry in any of its vectors.

In this work, the dual ROB is constructed using non-negative matrix factorization (NMF). Specifically,
$m_1$ solution snapshots of (\ref{eqn:fem1}), $\{\lambdabold_1^{(1)}, \dots, \lambdabold_1^{(m_1)}\}$,
are computed at scale $1$ for different prescribed boundary displacements and collected in the {\it dual}
snapshot matrix 
\begin{equation*}
  \Ybm_1^{\lambda} = \begin{bmatrix} \lambdabold_1^{(1)} & \cdots & \lambdabold_1^{(m_k)} \end{bmatrix}
\end{equation*}
and the dual ROB $\Wbm_1$ is constructed from the left factor of the NMF of $\Ybm_k^{\lambda}$
\cite{balajewicz2016projection}.

The dimensionality of the reduced governing equations (\ref{eqn:rom1}) is further reduced at scale $1$
by substituting (\ref{eqn:dualansatz}) and projecting the gap function $\gbm_1$ onto the column space
of $\Wbm_1$. Hyperreduction of the internal force is also performed. This leads to the PROM
\begin{subequations}
\begin{align*}
  \fbm_{1_r}^{int}\left(\bar\ybm_1\right) + \Vbm_1^T\Gbm_1\left(\overline\Vbm_1\bar\ybm_1\right)\Wbm_1\zbm_1 &= 0 \\
  \Wbm_1^T\gbm_1\left(\overline\Vbm_1\bar\ybm_1\right) &\ge 0 \\
  \zbm_1 &\le 0 \\
  \zbm_1^T \Wbm_1^T\gbm_1\left(\overline\Vbm_1\bar\ybm_1\right) &= 0
\end{align*}
\end{subequations}

Typically, the evaluations of the gap function and its Jacobian do not require a reconstruction of the
full state but only its restriction to the contact surface. Furthermore, Galerkin projection of the contact
force term can be optimized by accounting for the sparsity of the Jacobian. Specifically, only the row-wise
restriction of $\Gbm_1$ to the contact surface is non-zero. Nevertheless, these evaluations may still
incur a substantial computational cost. In principal, hyperreduction can be applied to further accelerate
the evaluation of the reduced gap function and its Jacobian. This is an active topic of research but
is not employed in the present work. However, it is noted that in the case of linear constraints, the proposed
reduction of the dual variables leads to terms involving reduced-order matrices that are \emph{precomputable}
and as such does not generate any bottleneck in the online solution of the reduced-order discrete microscale
equations. Consequently, just like in the case of any other linear terms, the efficient processing of such terms
does not require any hyperreduction.

\bibliography{pap}{}

\end{document}